\documentclass[fleqn,usenatbib]{mnras}

\usepackage{mathptmx}

\usepackage[T1]{fontenc}

\DeclareRobustCommand{\VAN}[3]{#2}
\let\VANthebibliography\thebibliography
\def\thebibliography{\DeclareRobustCommand{\VAN}[3]{##3}\VANthebibliography}

\usepackage{graphicx}	
\usepackage{amsmath}	
\usepackage{amssymb}	

\newcommand{\heii}{He\,\textsc{ii}}
\newcommand{\lya}{Ly$\alpha$}

\newcommand{\lum}{erg\,s$^{-1}$}
\newcommand{\sfr}{$M_\odot$\,yr$^{-1}$}

\title[VANDELS: X-ray binaries at high redshifts]{The VANDELS Survey: New constraints on the high-mass X-ray binary populations in normal star-forming galaxies at $\mathbf{3 < z < 5.5}$}

\author[A. Saxena et al.]{A. Saxena$^{1}$\thanks{E-mail: aayush.saxena@ucl.ac.uk},
R. S. Ellis$^{1}$,
P. U. F\"{o}rster$^{1}$,
A. Calabr\`{o}$^{2}$,
L. Pentericci$^{2}$,
A. C. Carnall$^3$,
M. Castellano$^{2}$,
\newauthor
F. Cullen$^3$,
A. Fontana$^2$,
M. Franco$^4$,
J. P. U. Fynbo$^5$,
A. Gargiulo$^6$,
B. Garilli$^6$,
N. P. Hathi$^7$,
\newauthor
D. J. McLeod$^3$,
R. Amor\'{i}n$^{8,9}$
and G. Zamorani$^{10}$
\\
$^{1}$Department of Physics and Astronomy, University College London, Gower Street, London WC1E 6BT, UK \\
$^{2}$INAF -- Osservatorio Astronomico di Roma, via Frascati 33, I-00078 Monteporzio Catone, Italy \\
$^{3}$SUPA -- Scottish Universities Physics Alliance, Institute for Astronomy, University of Edinburgh, Royal Observatory, Edinburgh EH9 3HJ \\
$^4$Centre for Astrophysics Research, Department of Physics, Astronomy and Mathematics, University of Hertfordshire, College Lane, Hatfield AL10 9AB, UK \\
$^5$Cosmic DAWN Center, Niels Bohr Institute, University of Copenhagen, Juliane Maries Vej 30, DK-2100 Copenhagen, Denmark \\
$^6$INAF -- Istituto di Astrofisica Spaziale e Fisica Cosmica Milano, via
A.Corti 12, 20133 Milano, Italy \\
$^7$Space Telescope Science Institute, 3700 San Martin Drive, Baltimore, MD 21218, USA \\
$^8$Instituto de Investigaci\'on Multidisciplinar en Ciencia y Tecnolog\'ia, Universidad de La Serena, Ra\'ul Bitr\'an, 1305 La Serena, Chile \\
$^9$Departamento de F\'isica y Astronom\'ia, Universidad de La Serena, Av. Juan Cisternas 1200 Norte, La Serena, Chile \\
$^{10}$INAF – OAS Bologna, Via P. Gobetti 93/3, 40129 Bologna, Italy
}

\date{Accepted 2021 May 26. Received 2021 May 25; in original form 2021 April 2.}

\pubyear{2021}

\begin{document}
\label{firstpage}
\pagerange{\pageref{firstpage}--\pageref{lastpage}}
\maketitle

\begin{abstract}
We use VANDELS spectroscopic data overlapping with the $\simeq$7 Ms {\it Chandra Deep Field South} survey to extend studies of high-mass X-ray binary systems (HMXBs) in 301 normal star-forming galaxies in the redshift range $3 < z < 5.5$. Our analysis evaluates correlations between X-ray luminosities ($L_X$), star formation rates (SFR) and stellar metallicities ($Z_\star$) to higher redshifts and over a wider range in galaxy properties than hitherto. Using a stacking analysis performed in bins of both redshift and SFR for sources with robust spectroscopic redshifts without AGN signatures, we find convincing evolutionary trends in the ratio $L_X$/SFR to the highest redshifts probed, with a stronger trend for galaxies with lower SFRs. Combining our data with published samples at lower redshift, the evolution of $L_X$/SFR to $z\simeq5$ proceeds as $(1 + z)^{1.03 \pm 0.02}$. Using stellar metallicities derived from photospheric absorption features in our spectroscopic data, we confirm indications at lower redshifts that $L_X$/SFR is stronger for metal-poor galaxies. We use semi-analytic models to show that metallicity dependence of $L_X$/SFR alone may not be sufficient to fully explain the observed redshift evolution of X-ray emission from HMXBs, particularly for galaxies with SFR $<30$\,\sfr. We speculate that reduced overall stellar ages and ``burstier'' star-formation histories in the early Universe may lead to higher $L_X$/SFR for the same metallicity. We then define the redshift-dependent contribution of HMXBs to the integrated X-ray luminosity density and, in comparison with models, find that the contribution of HMXBs to the cosmic X-ray background at $z>6$ may be $\gtrsim 0.25$ dex higher than previously estimated. 
\end{abstract}




\section{Introduction}


Understanding the physical processes that governed cosmic reionisation remains a fundamental issue in astronomy. This important transition in the nature of the intergalactic medium (IGM) is thought to have begun at a redshift $z \gtrsim 15$ (\citealt{bro11}) and completed by a redshift $z \approx 6$ \citep{fan06, bec13}. While it is frequently assumed the process was driven by star-forming galaxies \citep{rob13, rob15}, there are outstanding questions related to their photoionisation production rates and the extent to which Lyman continuum (LyC) photons can escape absorption in the interstellar and circumgalactic gas (see \citealt{star16} for a recent review).

Recent attention has focused on the possibility that reionisation ended rather abruptly as evidenced from estimates of the rapidly evolving neutral fraction in the IGM over $5.5 < z < 7.5$ \citep{nai20, ouc20}. This might suggest additional contributions from rarer sources with harder radiation fields such as active galactic nuclei (AGN). At present there is only limited and indirect evidence for AGN activity in $z>7$ galaxies, primarily through the presence of high ionisation emission lines in a few examples \citep{lap17, mai18}. However, the surprising presence of Lyman $\alpha$ emission in several luminous galaxies at redshifts where it should be resonantly scattered by a predominantly neutral IGM may provide further support for hard radiation fields \citep[e.g.][]{star17}.

A further topic that has merited attention, and represents the focus of this article, relates to the contribution of particularly the high mass X-ray binaries (HMXBs) within star-forming galaxies to the cosmic X-ray background at high redshifts. X-ray heating of the very early Universe is expected to play an important role in shaping reionisation both temporally and spatially \citep{war09, mes13, pac14, mei17, eid18} at scales between the H\,\textsc{ii} bubbles at galactic scales \citep[e.g][]{mad17}. The role of harder X-ray photons resulting from HMXBs has also been recently investigated to explain observations of high ionisation nebular emission lines such as \heii\ in star-forming galaxies, both in the local Universe \citep{sch19, sen20, keh21} and at high redshifts \citep{sax20}. 

HMXBs are binary stellar systems where a black hole or neutron star accretes material from a companion star that has a mass $\gtrsim10$\,M$_\odot$ \citep{tau06}. The resulting energetic feedback can increase the ionising photon output of the system and may further create channels for LyC radiation to escape \citep[e.g.][]{pla19}. To understand whether HMXBs can make a significant contribution to the photoionising budget required for reionising the Universe at $z>6$, it is necessary to understand the demographics of X-ray emission powered by HMXBs in star-forming galaxies at intermediate redshifts, as well as to constrain any evolution with redshift in the contribution of HMXBs to the overall ionising budget of galaxies. Some models of HMXBs \citep{fra13a, mad17} have indicated that HMXBs may provide a greater contribution than AGN to the global X-ray background at $z>5$ \citep{air15}.

Several studies have found strong correlations between the X-ray emission from HMXBs and the star-formation rate (SFR) of the host galaxy \citep{gri03, ran03, col04, hor05, leh10, leh16, leh19, kou20}. This correlation is readily understood given the short lifetimes of the high mass stellar companions, typically $\sim 10 \, \mathrm{Myr}$ \citep{ibe95, bod12, ant16}. Recent studies have further shown that the normalisation of the $L_X$-SFR relation increases with redshift over $0 < z < 2.5$ \citep{bas13a, leh16, air17, for19} with the most likely explanation for this evolution being an underlying dependence of $L_X$/SFR on metallicity \citep{lin10, fra13a, fra13b, mad17, for20, leh21}. Local low metallicity dwarf galaxies appear to host a larger number of HMXBs compared to more metal-rich galaxies in the local Universe \citep{kar11, pre13, dou15, bro16, kov20, for20}, and recent studies at intermediate redshift have confirmed a similar metallicity trend \citep{for19}. 

As HMXBs are likely driven by wind accretion \citep[see][for example]{kre21}, increased X-ray emission in binaries with a metal-deficient donor star may arise in part from the greater mass loss and hence, increased accretion onto the compact companion. Several theoretical studies have also suggested that the formation of more massive accretors, i.e. black holes or neutron stars at lower metallicities \citep{heg03} may also lead to elevated X-ray emission \citep[e.g.][]{lin10}. The exact drivers of the metallicity dependence of HMXB emission, however, are still not clear.

Most studies exploring the metallicity dependence of the X-ray luminosity of HMXBs have been limited to low redshifts, and studies exploring the redshift evolution of HMXB emission at high redshifts have typically relied on photometrically selected samples from deep fields overlapping with X-ray coverage with little to no metallicity information. Attempts have been made to model the metallicity dependence of HMXBs and predict their contribution to the cosmic X-ray background at high redshifts \citep[e.g.][]{mad17}, and to quantify the metallicity dependence of the X-ray luminosity functions due to HMXBs \citep[e.g.][]{leh21}. A crucial missing link may be provided by a study of both the metallicity dependence and the redshift evolution of the HMXB population at high redshifts using spectroscopic data.

In this work we attempt to quantify both the redshift evolution of HMXB emission and its dependence on metallicity by exploiting the overlap of the recently completed VANDELS redshift survey \citep{mcl18, pen18, gar21} with deep 7 Ms exposures of the \textit{Chandra Deep Field South} (CDFS, \citealt{cdfs7ms}). This unique combination of data sets enables more accurate measurements of the underlying stellar population properties of galaxies at $z>3$ compared to earlier studies that relied heavily on photometric redshifts. We probe the HMXB population in spectroscopically confirmed galaxies over the redshift range $3 < z < 5.5$ to significantly lower stellar masses and SFRs than possible before, and the availability of high quality spectra enables accurate derivation of the important galaxy physical parameters compared to previous studies that primarily relied on photometric redshifts alone. 

Thanks to VANDELS spectra we can now also extend the redshift range over which accurate constraints on HMXB emission can be obtained. Therefore, the fundamental goals of our paper are to (i) evaluate the evolutionary trends of X-ray emission from HMXBs, (ii) determine whether such evolution is largely the result of increasing metallicity with cosmic time, and (iii) consider the implications for the contribution of HMXBs to the cosmic X-ray background and to cosmic reionisation as a result. 


This paper is organised as follows: in \S\ref{sec:Data} we describe the VANDELS and CDFS {\it Chandra} X-ray data used in this study. In \S\ref{sec:X-ray photometry} we present our methodology for X-ray photometry and stacking. In \S\ref{sec:Properties} we present the physical properties of galaxies in our sample determined by constraining the spectral energy distributions using multi-wavelength data, and show the distribution of stacked X-ray counts, fluxes and luminosities in bins of redshifts and star-formation rates. In \S\ref{sec:Redshift evolution of L_X/SFR} we constrain the redshift evolution of X-ray luminosities from HMXBs. In \S\ref{sec:Stellar metallicity dependence of XRBs} we explore the dependence of X-ray emission on stellar metallicities, and test whether the cosmic evolution of metallicities can explain the observed redshift evolution of the HMXB output. In \S\ref{sec:XRB emission from strong Lyalpha emitters} we present the stacked X-ray properties of specifically those galaxies that also show strong Lyman alpha emission in their VANDELS spectra, and in \S\ref{sec:Discussion} we discuss the implications of the redshift and metallicity dependence of HMXB output on the cosmic X-ray background. We summarise the key findings of this study in \S\ref{sec:Conclusions}.

Throughout this paper, we assume a $\Lambda$CDM cosmology with $\Omega_m = 0.3$ and H$_0=67.7$ km\,s$^{-1}$\,Mpc$^{-1}$, taken from \citet{planck}. We use cgs units for flux and luminosity measurements and the AB magnitude system \citep{oke83}. All logarithms used in this study are logarithms to base 10.

\section{Data}
\label{sec:Data}
Since our main goal is to study the X-ray emission from HMXB populations in high redshift galaxies, we construct a sample of galaxies with high quality spectroscopic measurements that overlaps with the sensitive X-ray data in CDFS. These data sets are described in this section.

\subsection{VANDELS}
The spectroscopically selected galaxies used in this study are drawn from VANDELS, which is a recently completed public ESO survey of the Chandra Deep Field South (CDFS) and the UKIDSS Ultra Deep Survey (UDS) fields, which are part of the Cosmic Assembly Near-IR Deep Extragalactic Legacy Survey (CANDELS) \citep{gro11, koe11}, using the now decommissioned VIMOS on the Very Large Telescope (VLT). The survey description and initial target selection strategies can be found in \citet{mcl18} and data reduction and redshift determination methods can be found in \citet{pen18}. The details of the final public VANDELS data release can be found in \citet{gar21}. The data release contains spectra of approximately $2100$ galaxies in the redshift range $1.0<z<7.0$, with over $70\%$ of the targets having at least 40 hours of on-source integration time. VANDELS spectra have high signal-to-noise ratios (S/Ns) owing to deep exposure times, which ensures accurate spectroscopic redshift determination, in addition to the derivation of robust stellar metallicities and better constraints on physical parameters such as stellar masses and star-formation rates. 

In this work, we select VANDELS spectra of Lyman-break galaxies at $z>3$ in the CDFS field given the availability of the deepest available X-ray image taken by \textit{Chandra} in this field. These galaxies benefit from both \textit{Hubble Space Telescope (HST)} and ground-based photometric data. From the VANDELS database, we only select those sources that have a redshift reliability flag of either 3 or 4, which guarantees that the redshift measured by the VANDELS team has $>95$\% probability of being correct \citep[see][]{pen18}. We do not apply any further selection cuts other than $z>3$ and the requirement of the above mentioned redshift reliability flags.

\subsection{The \textit{Chandra} Deep Field South (CDFS) 7 Ms image}
\begin{figure}
	\centering
	\includegraphics[scale=0.24]{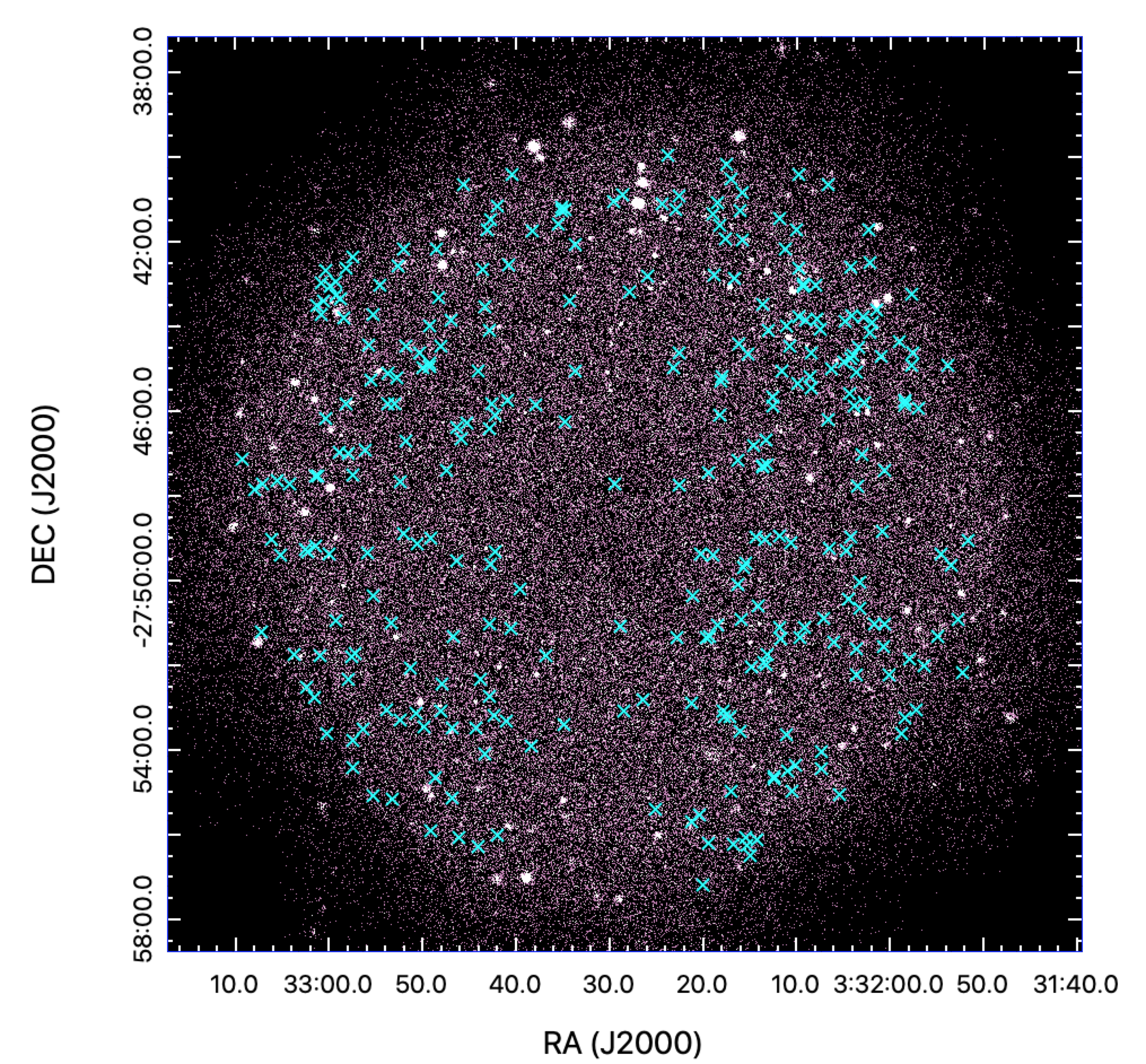}
	\caption{Sky distribution of 318 VANDELS sources at $z>3$ with reliable spectroscopic redshifts used in this study (cyan crosses). The CDFS 7 Ms image in the $0.8-3$ keV  medium band from \citet{gia19} is shown in the background. We only consider sources that lie within a radius of 9 arcminutes from the centre of the CDFS image to ensure relatively well behaved PSFs and accurate background estimation.}
	\label{fig:ra_dec}
\end{figure}

The X-ray data used in this study is the deepest available X-ray image, taken using the \emph{Chandra X-ray Observatory} in CDFS, with a total of 7 Ms of exposure time covering an area of $\sim 485$ arcmin$^2$ collected over a period of more than a decade \citep{cdfs7ms}\footnote{The images and catalogues are publicly available at \url{http://personal.psu.edu/wnb3/cdfs/cdfs-chandra.html}}. Additional data products in the CDFS include the effective exposure map and the PSF map, which we will use for aperture photometry. A detailed description of the data reduction methods and final data products in the CDFS field that have been used in this study can be found in \citet{gia19}.  

In this study we use the $0.8-3$ keV (medium band) \textit{Chandra} image (similar to \citealt{sax20b}) for two reasons. First, as demonstrated by \citet{gia19}, using the $0.8-3$ keV image instead of the standard soft X-ray $0.5-2$ keV \textit{Chandra} band, leads to a higher number of counts recovered from faint objects, due to the higher transmissivity of the $0.8-3$ keV band. Second, owing to the redshift distribution of sources in this study, the $0.8-3$ keV band comes closest to rest-frame energies in the range $2-10$ keV. Therefore, the uncertainties arising from the application of $k$-corrections are minimised by using the medium band image.

\subsection{Final sample}
To achieve a balance between the increasing degradation of the point spread function (PSF) away from the pointing centre of the \textit{Chandra} 7 Ms image, which is a combination of images with different pointing centres and roll angles \citep{cdfs7ms}, and the need for high number statistics, we consider galaxies that lie within a radial distance of 9 arcminutes from the pointing centre of the CDFS image. This results in the best compromise between the total number of spectroscopically confirmed galaxies within the X-ray footprint and adequate X-ray coverage for reliable background estimation for sources.

We then identify and remove possible AGN in our sample by cross-matching the coordinates of our galaxies taken from the CANDELS catalogues \citep{guo13} with the CDFS 7 Ms X-ray catalogue \citep{cdfs7ms} using a conservative matching radius of 4 arcseconds. Specifically at $z>3$, the overwhelming majority of detections in the \citet{cdfs7ms} catalogue are identified as AGN and have high X-ray luminosities, i.e. $L_X \gtrsim 10^{42}$\,\lum\ (see also \citealt{mag20} and \S\ref{sec:agn_removal}). A conservative matching radius also ensures that contamination from bright X-ray sources in the vicinity of our galaxies is not mistakenly included in the aperture photometry that will follow. 

Therefore, after removing possible AGN the total number of spectroscopically confirmed sources with reliable redshifts at $z>3$ that lie within the X-ray footprint in CDFS is 318 and the spatial distribution of these sources is shown in Figure \ref{fig:ra_dec}.

\section{X-ray photometry}
\label{sec:X-ray photometry}

\subsection{Aperture photometry}
We follow the methodology described by \citet{sax20b} to estimate X-ray fluxes for the 318 star-forming galaxies within the CDFS footprint in the following way. We perform aperture photometry to measure X-ray fluxes (using \textsc{photutils}; \citealt{photutils}) at the Right Ascension (RA) and Declination (Dec) of each source taken from the CANDELS source catalogues. The X-ray counts from the source are measured by placing a circular aperture with a size that is 90\% of the size of the effective \textit{Chandra} PSF at the position of each galaxy. 

The local background is measured by placing a circular annulus with an inner radius that is $5''$ larger than the aperture radius, and an outer radius that is $10''$ larger than the aperture radius, centred on the same position as the circular aperture. Within the annulus, we mask pixels that are brighter than $4\sigma$ to calculate the background, similar to \citet{sax20b}. The total number of counts from the source ($C_\textrm{gal}$) within the area encompassed by the circular aperture ($A_\textrm{gal}$), and the background ($C_\textrm{bkg}$) measured within the area encompassed by the annulus ($A_\textrm{bkg}$) are recorded, in addition to the effective exposure times for each area, $t_\textrm{gal}$ and $t_\textrm{bkg}$ respectively. We then follow \citet{for19} and \citet{sax20b} and calculate the background subtracted counts as: 
\begin{equation}
	C_\textrm{bkgsub} = C_\textrm{gal} - C_\textrm{bkg} \times \left(\frac{A_\textrm{gal}\times t_\textrm{gal}}{A_\textrm{bkg}\times t_\textrm{bkg}}\right)
\end{equation}

Since the counts from individual galaxies at these redshifts are expected to be very low, we apply the \citet{geh86} approximation to establish confidence limits on the measured counts in the circular aperture.

To convert background subtracted counts in the $0.8-3$ keV band to X-ray fluxes in the standard $2-10$ keV band, we follow the procedure outlined in \citet{sax20b} and assume a spectral model to calculate the effective photon energy ($E_\textrm{eff}$) in the band and the appropriate $k$-correction ($k_\textrm{corr}$). We assume a model with an un-obscured power-law spectrum with photon index $\Gamma=2.0$ \citep[e.g.][]{bro16} and a galactic extinction value of $5\times 10^{20}$ cm$^{-2}$ \citep{voo12}, which is the average value observed for star-forming galaxies at high redshifts in cosmological simulations. Note here that we do not consider any redshift dependence of the galactic extinction. We then use \textsc{pimms}\footnote{\url{https://heasarc.gsfc.nasa.gov/docs/software/tools/pimms.html}} to calculate the effective photon energy, $E_\textrm{eff}$, required to convert observed counts in the $0.8-3$ keV band to fluxes in the observed $2-10$ keV band:
\begin{equation}
	F_{\textrm{2-10 keV}} = \frac{C_\textrm{bkgsub}}{t_\textrm{gal}} \times E_{\textrm{eff}}
\end{equation}

To calculate rest-frame X-ray luminosities in the $2-10$ keV band, we use luminosity distances, $D_L$, determined from the reliable spectroscopic redshifts of our VANDELS galaxies and apply the $k$-correction, $k_\textrm{corr} = (1+z)^{\Gamma-2.0}$:
\begin{equation}
	L_{\textrm{2-10 keV}} = F_{\textrm{2-10 keV}} \times 4\pi D_L^2 k_\textrm{corr}
\end{equation}

\subsection{Identifying and removing possible AGN}
\label{sec:agn_removal}
Out of 318 sources for which aperture photometry was performed, we identified 17 sources with $L_X > 10^{42}$ erg\,s$^{-1}$, which is commonly used as the limit above which X-ray emission from the source can be attributed to AGN \citep[see][for example]{mag20}. Upon visual inspection, all these sources were consistent with either being bright themselves or lying adjacent to a bright point source that was not in the \citet{cdfs7ms} source catalogue. To safeguard against any biases introduced by the inclusion of AGN in our sample, we identify all 17 of these sources as possible AGN and remove them from the analysis that follows.

We note here that it may be possible that rare star-forming galaxies with unusually bright X-ray emission may be excluded from the sample by introducing the cut above. However, the fraction of non-AGN sources with $L_X \gtrsim 10^{42}$\,\lum\ at $z>3$ is extremely low. For example, only two sources with $L_X \gtrsim 10^{42}$\,\lum\ are classified as galaxies at $z>3$ in the \citet{cdfs7ms} CDFS X-ray catalogue, with the large majority of such sources being classified as AGN. A more robust characterisation of sources lying at the $L_X = 10^{42}$\,\lum\ limit at high redshifts may be possible using the BPT \citep{bal81} classification based on rest-frame optical emission lines with the \emph{JWST} in the near future. However, since this study is geared towards exclusively studying the X-ray emission from galaxies, we take the conservative approach of classifying all sources with $L_X \gtrsim 10^{42}$\,\lum\ as possible AGN.

Even with the cut in X-ray luminosities, there may still be obscured AGN in our sample. However, \citet{vit18} have shown that the fraction of obscured AGN drops significantly at luminosities below $10^{43}$ \lum\ at $z>3$, which should minimise the chances of contamination in our $L_X < 10^{42}$ \lum\ sample. Further, using the CDFS 7 Ms data, \citet{cir19} found that even obscured AGN at $z>2.5$ can have X-ray luminosities in excess of $10^{44}$ \lum\ which would be excluded from our sample. Finally, we do not see any strong AGN emission lines (such as N\,\textsc{v} $\lambda 1240$) in the VANDELS spectra of the remaining galaxies in our sample. Therefore, we conclude that the likelihood of AGN contamination after removing sources with $L_X > 10^{42}$ \lum\ is minimal, and the total number of star-forming galaxies at $z>3$ in our final catalogue, excluding these possible AGN, is 301.

In the following section we measure the physical properties of the galaxies in the final sample, which are also used to create bins within which the individual X-ray measurements for our galaxy sample can be stacked.

\section{Host galaxy physical properties}
\label{sec:Properties}

\begin{figure*}
    \centering
    \includegraphics[width=\textwidth]{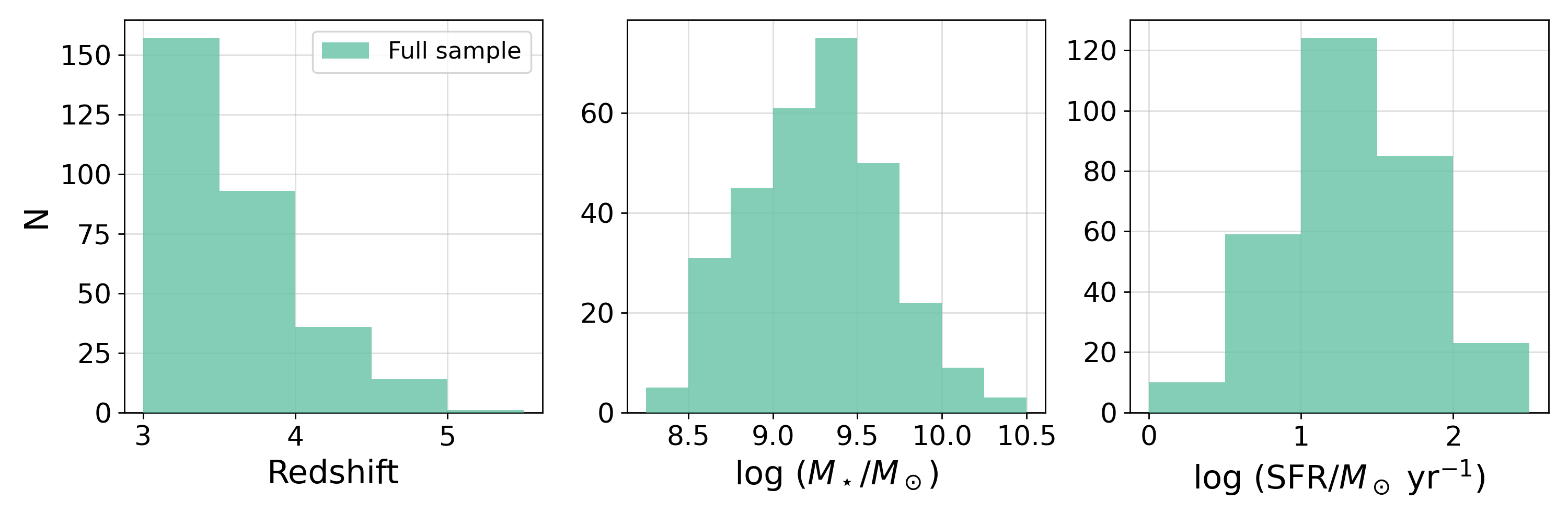}
    \caption{Distribution of redshift ($z$; \textit{left}), stellar masses ($M_\star$; \textit{middle}) and UV-corrected star-formation rates (SFR; \textit{right}) for 301 VANDELS galaxies in the final galaxy sample considered in this study. These galaxies sample the Lyman-break galaxy (LBG) population at $3 < z < 5.5$.}
    \label{fig:sample_properties}
\end{figure*}

Given the unique combination of accurate spectroscopic redshifts and excellent photometric data available from space-based and ground-based telescopes ranging from ultraviolet (UV) to mid-infrared (MIR) wavelengths in the CDFS field, we can derive reliable physical property measurements for all the VANDELS galaxies selected in this study. We use the PSF-homogenised photometric catalogues created by the VANDELS team and presented in \citet{mcl18} to obtain best-fit spectral energy distribution (SEDs) and derive properties such as stellar masses ($M_\star$), dust attenuation ($A_V$), star-formation rates (SFRs), and rest-frame absolute UV magnitudes ($M_{\textrm{UV}}$) for the 301 star-forming galaxies in our final sample using \textsc{Bagpipes} \citep{car18}. 

The SED fits were performed using $Z=0.2$\,$Z_\odot$ metallicity versions of the updated \citet{bc03} models \citep[see][]{che16} using the MILES spectral library \citep{fal11}, with a \citet{kro01} initial mass function (IMF). Among the metallicities available in the BC03 models, $0.2\,Z_\odot$ is the most appropriate value for the redshift range studied in this work \citep[e.g.][]{cul19}. The redshifts for the SED fitting were fixed to the VANDELS spectroscopic redshifts \citep[see][]{pen18}. The derived star-formation rates are corrected taking into account dust attenuation, $A_V$, adopting the \citet{cal00} dust attenuation law. The rest-frame magnitudes were calculated using a 200\,\AA\ wide top-hat filter centred at 1500\,\AA. For full details of the SED fitting techniques, model assumptions, and derived physical parameters we refer the readers to \citet{mcl18}. The knowledge of spectroscopic redshifts has been shown to provide considerably more reliable measurements of galaxies properties such as stellar masses and star-formation rates, compared to some earlier samples based upon photometric redshifts \citep[see][for example]{bun05}.

\subsection{Stellar masses, star formation rates and binning}
\label{sec:bins}
The distribution of spectroscopic redshifts, stellar masses and star-formation rates for galaxies in this study are shown in Figure \ref{fig:sample_properties}. The stellar masses of galaxies in our sample range from $\log(M_\star/M_\odot) = 8.3 - 10.5$, with a median stellar mass of $\log(M_\star/M_\odot) = 9.3$. The dust-corrected star-formation rates for our galaxies range from $1-270$ \sfr, with a median SFR of 20 \sfr. The specific star-formation rate (sSFR), which is the star-formation rate per unit stellar mass, ranges from $\log(\rm{sSFR/yr}^{-1}) = -8.1$ to $-7.7$ for our sample.

\begin{figure}
	\centering
	\includegraphics[scale=0.4]{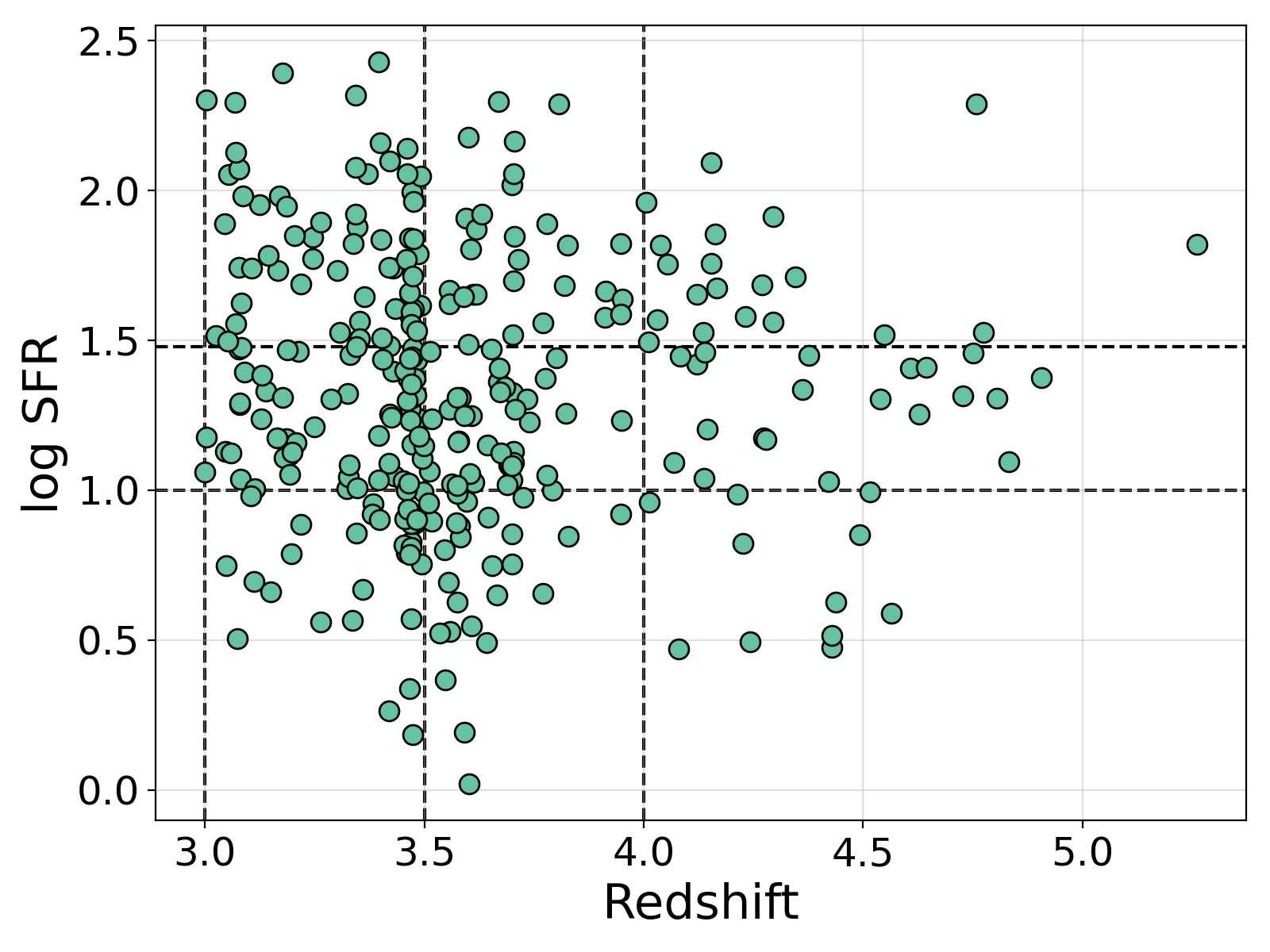}
	\caption{Distribution of UV-corrected SFR with redshift for our final sample of galaxies. Our sample probes fainter SFRs at higher redshifts compared to previous studies probing X-ray emission from galaxies at high redshifts. The dashed lines indicate the choice of bins in both SFR and redshift.}
	\label{fig:sfr_redshift}
\end{figure}

Since X-ray emission powered by HMXBs in star-forming galaxies is expected to be very faint at $z>3$, stacking is needed to derive meaningful signals to study the dependence of X-ray emission on galaxy properties. Emission from HMXBs is expected to correlate with SFRs, therefore separating galaxies in SFR bins will help probe the evolution of X-ray emission over redshifts in a way that is not affected significantly by incompleteness. Therefore, we have binned our galaxies in the SFR-redshift parameter space, as indicated by dashed lines in Figure \ref{fig:sfr_redshift}. The redshift bins are:

(1) $3 \le z < 3.5$

(2) $3.5 \le z < 4$

(3) $z \ge 4$

Within each redshift bin, we also bin in star-formation rates:

(1) SFR $< 10$ \sfr

(2) $10 \leq$ SFR (\sfr) $< 30$

(3) SFR $\geq 30$ \sfr

Having access to rest-frame UV spectra from VANDELS also enables us to obtain stacked spectra of galaxies within these bins to enable accurate metallicity measurements, the methodology for which is highlighted in the following sub-section. We show the number of sources and median physical properties of galaxies in these bins in Table \ref{tab:subset_props}.

\begin{table*}
    \centering
    \begin{minipage}{0.8\linewidth}\centering
    \caption{Median redshift and physical properties of galaxies in bins of star-formation rate and redshift.}
    \begin{tabular}{l c c c c c c c}
    \hline 
    Subset & ID & N & Redshift & SFR (\sfr) & $\log(M_\star/M_\odot)$ & $\log$(sSFR/yr$^{-1}$) & $\log(Z_\star/Z_\odot)$ \\
    (1) & (2) & (3) & (4) & (5) & (6) & (7) & (8) \\
    \hline \hline
    
    $\mathbf{3 \le z < 3.5}$ & \textbf{1} \\
    SFR < 10 \sfr & 11 & 31 & 3.45 & 6.4 & 8.9 & $-8.09$ & $-1.01 \pm 0.44$   \\
    $10 \leq$ SFR (\sfr) $< 30$ & 12 & 63 & 3.35 & 17.2 & 9.1 & $-7.86$ & $-0.66 \pm 0.14$ \\
    SFR $\geq 30$ \sfr & 13 & 63 & 3.35 & 61.3 & 9.5 & $-7.71$ & $-0.79 \pm 0.13$ \\
    \hline
    
    $\mathbf{3.5 \le z < 4}$ & \textbf{2} \\
    SFR < 10 \sfr & 21 & 27 & 3.60 & 6.3 & 9.0 & $-8.20$ & $-1.93 \pm 0.62$ \\
    $10 \leq$ SFR (\sfr) $< 30$ & 22 & 37 & 3.67 & 17.1 & 9.0 & $-7.97$ & $-1.40 \pm 0.43$ \\
    SFR $\geq 30$ \sfr & 23 & 29 & 3.70 & 58.8 & 9.5 & $-7.73$ & $-0.54 \pm 0.22$ \\
    \hline
    
    $\mathbf{z > 4}$ & \textbf{3} \\
    SFR < 10 \sfr & 31 & 11 & 4.43 & 4.2 & 9.1 & $-8.48$ & $-$ \\
    $10 \leq$ SFR (\sfr) $< 30$ & 32 & 20 & 4.40 & 20.4 & 9.1 & $-7.84$ & $-$ \\
    SFR $\geq 30$ \sfr & 33 & 20 & 4.17 & 50.0 & 9.5 & $-7.80$ & $-1.19 \pm 1.36$ \\
    \hline

    \end{tabular} \\
    \textit{Notes.} (1): Subsets in redshift and SFR, (2): Subset ID, (3): Number of sources in each subset, (4): Median redshift of galaxies in each subset, (5): Dust-corrected star-formation rate, (6): log stellar mass in units of solar mass, (7): log specific star-formation rate, (8): log stellar metallicity in units of solar metallicity.
    \label{tab:subset_props}
    \end{minipage}
\end{table*}

\subsection{Stellar metallicities of stacked UV spectra}
\label{sec:metallicity}
To estimate accurate stellar metallicities using rest-frame UV spectra from VANDELS, we stack the spectra of galaxies in each SFR-redshift bin (that were presented above). The stacking methodology we adopt closely follows that of \citet{mar18} and \citet{sax20}. Very briefly, each spectrum is first de-redshifted using the VANDELS spectroscopic redshifts \citep{pen18}. The rest-frame spectra are normalised using the mean flux density in the $1460-1540$\,\AA\ wavelength range, with a weight being assigned to the spectrum based on the average S/N in this range. The spectra are then resampled to a wavelength grid corresponding to the rest-frame wavelength coverage for each redshift bin. Sky residuals are masked and the spectra in each SFR-redshift bin are co-added using a weighted average method, where the weights are assigned as 1/err$^2$ (totalling to 1), where err is the error calculated in the rest-frame wavelength range $1460-1520$\,\AA. The errors on the stacked spectra are calculated using standard propagation equations, following the prescription outlined by \citet{gua17}.

We then estimate stellar metallicities ($Z_\star$) in terms of solar metallicity ($Z_\odot$) from the stacked spectrum in each SFR-redshift bin following the method of \citet{cal21}. A briefly summary of the method is given below, but we direct the readers to \citet{cal21} for a full description of the method. Very briefly, stellar metallicities are estimated by comparing the observer stellar photospheric absorption features at around $1501$\,\AA\ produced by ionised S\,\textsc{v} species in the photospheres of young and hot stars, and at around $1719$\,\AA\ produced by a complex of N\,\textsc{iv} $\lambda1718.6$, Si\,\textsc{iv} $\lambda \lambda 1772.5, 1727.4$ and multiple transitions of Al\,\textsc{ii} and Fe\,\textsc{iv} that range between 1705 and 1729\,\AA\ with the Starburst99 \citep{lei10} stellar synthesis models. As \citet{cal21} note, the $\lambda$1501 and $\lambda$1719\,\AA\ indices are largely unaffected by stellar age, dust attenuation, choice of IMF and nebular continuum emission or absorption. The solar metallicity in this method is assumed to be $Z_\odot = 0.0142$ \citep{asp09}, which is slightly different than the solar metallicity of 0.02 assumed in the BC03 models. 

We also note that the stellar metallicities measured for VANDELS galaxies by \citet{cal21} are highly comparable to a slightly different method implemented by \citet{cul19}. Further, \citet{cal21} found a very marginal difference between the metallicities measured using Starburst99 and models that include binary stars, such as BPASS \citep{eld17}, where this difference is often smaller than the uncertainties in the metallicity measurements themselves. The measured stellar metallicities and associated errors for stacked spectra in bins of redshift and SFR are also shown in Table \ref{tab:subset_props}. 

A limitation of comparing the stellar absorption features with Starburst99 models is that the lowest elemental abundances adopted in the model are $0.05\,Z_\odot$, which means that observational measurements that are indicative of seemingly lower metallicities than this limit must be extrapolated using calibrations derived at higher stellar metallicities. These extrapolations introduce additional uncertainties on measurements at the lowest stellar metallicities. We use the calibration derived by \citet{cal21} in these cases, and find that that metallicities below the model limits are always consistent within one standard deviation of the lowest $Z_\star$ Starburst99 models.

\subsection{Stacked X-ray measurements in redshift-SFR bins}
\begin{table*}
    \centering
    \begin{minipage}{0.7\linewidth}\centering
    \caption{Stacked X-ray properties of galaxy subsets used in this study.}
    \begin{tabular}{l c c c c c c}
    \hline 
    Subset & ID & N & Redshift & Net counts & $L_X$ & $\log(L_X$/SFR)  \\
           &    &   &          &            & ($\times10^{41}$ \lum)  & (\lum/$M_\odot$\,yr$^{-1}$) \\
    (1) & (2) & (3) & (4) & (5) & (6) & (7) \\
    \hline \hline
    
    $\mathbf{3 \leq z < 3.5}$ & \textbf{1} \\
    SFR < 10 \sfr & 11 & 31 & 3.45 & $<69.2$ & $<3.4$ & $<40.7$\\
    10 $\leq$ SFR (\sfr) < 30 & 12 & 63 & 3.35 & 57.9 $\pm$ 16.8 & 2.7 $\pm$ 1.1 & 40.2 $\pm$ 0.2 \\
    SFR $\geq$ 30 \sfr & 13 & 63 & 3.35 & 64.3 $\pm$ 16.4 & 2.9 $\pm$ 1.1 & 39.7 $\pm$ 0.3 \\
    \hline
    
    $\mathbf{3.5 \leq z < 4}$ & \textbf{2} \\
    SFR < 10 \sfr & 21 & 27 & 3.60 & $<44.5$ & $<3.9$ & $<40.9$ \\
    10 $\leq$~SFR~(\sfr)~$<30$ & 22 & 37 & 3.67 & 52.3 $\pm$ 15.0 & 3.9 $\pm$ 0.9 & 40.4 $\pm$ 0.2 \\
    SFR~$\geq 30$~\sfr & 23 & 29 & 3.70 & 61.2 $\pm$ 15.1 & 4.4 $\pm$ 1.4 & 39.9 $\pm$ 0.2\\
    \hline
    
    $\mathbf{z > 4}$ & \textbf{3} \\
    SFR~$<10$~\sfr & 31 & 11 & 4.43 & $<64.2$ & $<6.8$ & $<41.1$ \\
    $10 \leq$~SFR (\sfr)~$<30$ & 32 & 20 & 4.40 & 52.6 $\pm$ 19.4 & 6.6 $\pm$ 1.3 & 40.5 $\pm$ 0.2 \\
    SFR~$\geq 30$~\sfr & 33 & 20 & 4.17 & 62.2 $\pm$ 29.1 & 6.1 $\pm$ 2.1 & 40.0 $\pm$ 0.2 \\
    \hline

    \end{tabular} \\
    \textit{Notes.} (1)--(4): Same as Table \ref{tab:subset_props}, (5): Stacked background-subtracted X-ray counts, (6): stacked X-ray luminosity, (7): stacked X-ray luminosity per star-formation rate.
    \label{tab:subsets_xray}
    \end{minipage}
\end{table*}

As mentioned earlier, the S/N of individual X-ray measurements from star-forming galaxies at high redshifts are expected to be low. Therefore, we must rely on stacking the observed X-ray counts to boost S/N. Within the SFR and redshift bins presented in \S\ref{sec:bins} (see also Figure \ref{fig:sfr_redshift}), we now obtain the stacked X-ray luminosity of $N$ sources by calculating the weighted average of the X-ray luminosities ($L_{X,i}$) within a bin, where the weight ($w_{X,i}$) is the inverse square of the uncertainty in the corresponding X-ray fluxes ($\Delta F_{X,i}$), such that sources with higher uncertainties are assigned lower weights, and the sum of all weights is normalised to 1:
\begin{equation}
	L_{X}^{\textrm{stack}} = \sum\limits_{i}^N F_{X,i} . w_{X,i} \times 4\pi D_{L,i}^2 k_\textrm{corr}
\end{equation}
where $D_{L}$ is the luminosity distance of a galaxy calculated using its spectroscopic redshift. The errors on the stacked luminosity are calculated using standard propagation of uncertainty formulae.

The stacked $L_X$/SFR for each bin is calculated in a similar fashion, where the weighted average of all individual $L_X$/SFR values is calculated. We find that the relative errors on the SFR are negligible compared to the relative errors on $L_X$ (roughly a factor of 4 smaller) and therefore, we ignore the uncertainty on SFR when calculating the errors on stacked $L_X$/SFR for each bin.

For the lowest SFR bins (SFR $<10$\,\sfr) across redshifts, we do not find statistically significant X-ray counts even through stacking owing to low intrinsic X-ray luminosities as well as low number statistics. Therefore, we place 3 sigma upper limits on the stacked X-ray measurements for galaxies in the lowest SFR bin across redshifts. For the intermediate SFR bins ($10 \leq$~SFR\,(\sfr)~$< 30$), we find stacked X-ray luminosities of 2.7, 3.9 and 6.6 $\times 10^{41}$\,\lum\ in the first, second and third redshift bin, respectively. For the highest SFR bins ($\rm{SFR} \geq 30$\,\sfr), we find stacked X-ray luminosities of 2.9, 4.4 and 6.1 $\times 10^{41}$\,\lum\ for the first, second and third redshift bins, respectively. The stacked background-subtracted X-ray counts, luminosities and X-ray luminosity per SFR measurements for the redshift and SFR bins used in this study are shown in Table \ref{tab:subsets_xray}.

Having measured the physical properties of galaxies in our sample, created bins in redshift and SFR, and measured the stacked X-ray luminosities per SFR of galaxies in these bins, in the following sections we turn our attention to the observed trends of the X-ray output of HMXBs with redshift and metallicity.

\section{Redshift evolution of HMXB emission}
\label{sec:Redshift evolution of L_X/SFR}

In this section we use our new constraints on $L_X$/SFR at $z>3$ to determine the redshift evolution of the X-ray emission from HMXBs. We combine our measurements with observations in the literature of spectroscopically confirmed galaxies at lower redshifts to increase the redshift baseline, providing a more accurate normalisation of the redshift dependence of $L_X$/SFR. Following \citet{bas13a}, the redshift ($z$) dependence of $L_X$ and SFR can be parameterised as:
\begin{equation}
    \log(L_X) = A + B\log(\textrm{SFR}) + C\log(1+z)
\end{equation}

To constrain the redshift evolution over a range of redshifts, we anchor our new $z>3$ measurements using $L_X$/SFR measurements for analogues of Lyman-break galaxies from \citet{bro16} at $z<0.2$, and stacked measurements from the redshift bins of \citet{for20} at $z<1$ and \citet{for19} at $z\sim2$. We then derive the redshift dependence, including the lower redshift measurements, separately in the three SFR bins considered in this study to avoid complications introduced by incompleteness. We note here that the SFRs in the \citet{bro16} sample are estimated through SED fitting using UV+IR broadband data, similar to our study, whereas SFRs in the \citet{for19} and \citet{for20} studies are measured using both SED-fitting and H$\alpha$ measurements.

\begin{figure*}
    \centering
    \includegraphics[width=0.5\textwidth]{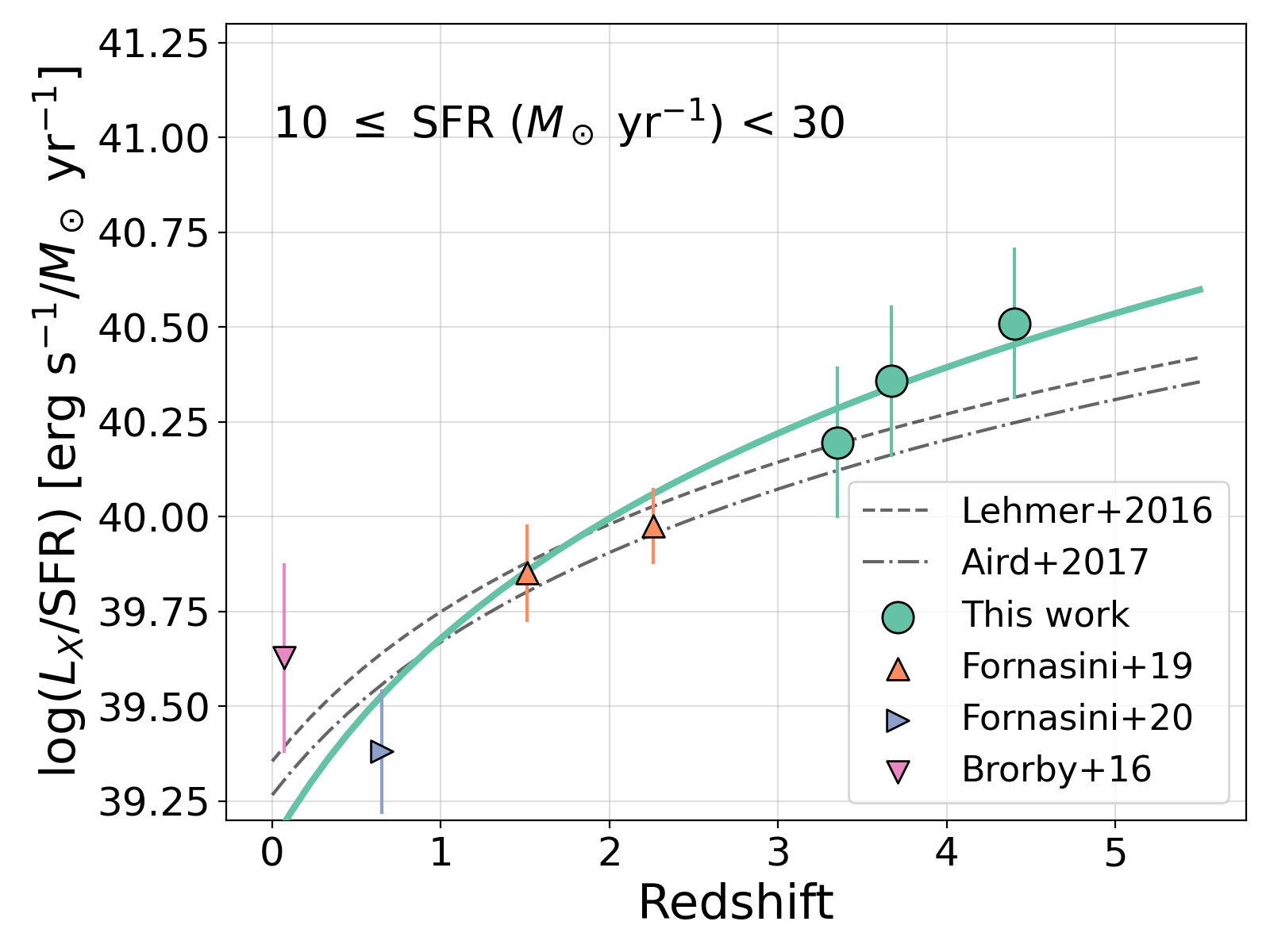}
    
    \includegraphics[width=0.9\textwidth]{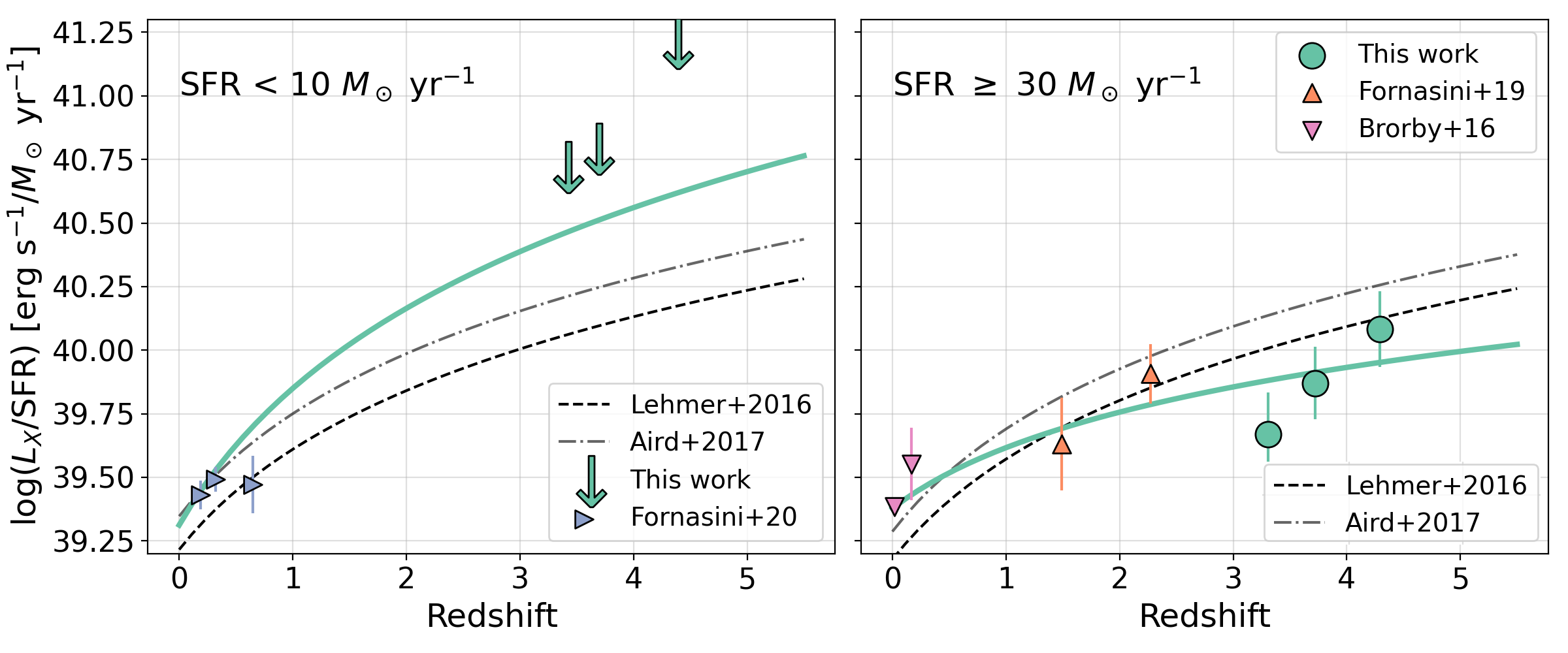}
    \caption{Redshift evolution of $L_X$/SFR for galaxies with $10 \leq$~SFR (\sfr)~$<30$ which has the best statistics (top), those with SFR $<10$ \sfr\ (bottom-left), and SFR~$\geq 30$~\sfr\ (bottom-right). Also shown are measurements from other spectroscopically confirmed samples of galaxies with similar properties in the literature: \citet{bro16} and \citet{for20} at $z<1$, and \citet{for19} at $z\sim2$. The best-fit relation derived between $L_X$/SFR and redshift by adding our measurements to these literature measurements for each SFR bin is shown as a green solid line. For the SFR $<10$ \sfr\ bin (bottom-left), we show the best-fit relation from the $10\leq$~SFR(\sfr)$<30$ bin (top). For comparison, we also show the best-fit relation from \citet{leh16} (dashed line) and \citet{air17} (dot-dashed line). We report stronger redshift evolution in the $10 \leq$~SFR (\sfr)~$<30$ bin, which is is also consistent with the limits we obtain in the SFR~$<10$~\sfr\ bin. We find weaker redshift evolution in the SFR~$\geq 30$~\sfr\ bin compared to earlier studies.} 
    \label{fig:lx_sfr_z_evol}
\end{figure*}

In Figure \ref{fig:lx_sfr_z_evol} we show the measured $L_X$/SFR for our $z>3$ galaxies along with measurements at lower redshift from the literature in the three SFR bins considered. We note that for $z>3$ galaxies in the SFR $<10$ \sfr\ bin, even the \textit{Chandra} 7 Ms image is not deep enough to obtain meaningful detections with stacking. The detection of X-ray emission due to HMXBs in low-mass galaxies at high redshifts has been historically challenging owing to the very low expected X-ray signal in such galaxies \citep[see][for example]{leh16}.

For the $10 \leq$~SFR~$<30$~\sfr\ bin, we obtain the following best-fit coefficients: $A = 39.22 \pm 0.4$, $B=0.94 \pm 0.28$ and $C = 1.78 \pm 0.09$, shown as a solid green line in Figure \ref{fig:lx_sfr_z_evol}, top panel. The value of the redshift evolution coefficient ($C$) we derive is larger than $C=1.31$ reported by \citet{leh16}, and $C=1.34$ reported by \citet{air17} in the $2-10$ keV X-ray band, shown as dashed and dot-dashed lines in the figure, respectively. In particular, our new measurements at $z>3.5$ lie above the predictions from previous best-fit relations. The redshift dependence we find in this SFR bin is also stronger than that reported by \citet{bas13a} with $C=0.89$. We note, however, that the \citet{bas13a} and \citet{leh16} studies relied only on photometric redshifts, whereas the \citet{air17} study had spectroscopic redshifts for roughly 10\% of their sources. The best-fit relation that we obtain here is solely from galaxies with high quality spectroscopic redshifts.

As mentioned earlier, with our $z>3$ data we could only place upper limits on $L_X$/SFR using stacking in the lowest SFR bin with SFR $<10$ \sfr. The upper limits we derive are compatible with the redshift evolution measured in the $10 \leq$~SFR (\sfr)~$<30$ bin, which is shown as a solid green line in the bottom left panel of Figure \ref{fig:lx_sfr_z_evol}, but also with predictions from \citet{leh16} and \citet{air17}. In the highest SFR bin containing galaxies with SFR~$\geq 30$~\sfr, however, we find the best-fit coefficients $A=39.93 \pm 0.96$, $B=0.68 \pm 0.32$ and $C=0.79 \pm 0.02$, shown as solid green line in the bottom-right panel of Figure \ref{fig:lx_sfr_z_evol}, which indicates weaker redshift evolution compared to our intermediate SFR bin as well as that reported by \citet{leh16} and \citet{air17}.

When we combine our measurements with those in the literature across all SFR bins, we find coefficients $A=40.03 \pm 0.16$, $B=0.62 \pm 0.05$ and $C=1.03 \pm 0.02$, which suggests a weaker redshift evolution in the global sample compared to only those galaxies with $10 \leq$~SFR~$<30$\,\sfr, and more in line with measurements from \citet{bas13a}, \citet{leh16} and \citet{air17}. Therefore, the emerging picture from our analysis of the redshift evolution trends in different SFR bins is that X-ray emission from low to intermediate star-formation rate galaxies evolves more strongly with redshift compared to the highly star-forming galaxies, even out to $z>3$. 

Such differential redshift trends in SFR bins would be expected if the evolution of $L_X$/SFR is driven primarily by evolving metallicities \citep{fra13a, bas13b, dou15, bro16, mad17, leh19, for20, leh21}. Low-mass, low-metallicity galaxies become increasingly dominant at higher redshifts, which may explain the stronger redshift evolution seen in the lower SFR bins in our study based on the star-forming galaxy main sequence and the fundamental mass-metallicity relations \citep[e.g.][]{cul19, cal21}. Such a metallicity dependence would in turn lower the expected $L_X$/SFR from galaxies with relatively high star-formation rates at high redshifts, as they would become metal-enriched soon after their formation. In the following section, we explore the dependence of $L_X$/SFR on stellar metallicity across redshifts in our sample, and investigate whether this metallicity dependence possibly evolves with redshift. 

\section{Metallicity dependence of HMXB emission}
\label{sec:Stellar metallicity dependence of XRBs}

It is important to note that the majority of $L_X$/SFR-metallicity constraints have been derived using measurements in the low redshift ($z<1$) Universe. Constraining the metallicity dependence of HMXBs out to the lowest metallicities is crucial to characterise the X-ray output of HMXBs from analogues of galaxies that are expected to be the key drivers of reionisation in the early Universe. We note here that the stellar metallicities of VANDELS galaxies at $z>3$ are more than an order of magnitude lower than those measured for HMXB populations at low redshifts, providing crucial constraints at the lowest metallicity regime at $z>3$ for the first time.

\subsection{Relation between $\mathbf{L_X}$/SFR and stellar metallicity}

\begin{figure}
    \centering
    \includegraphics[scale=0.4]{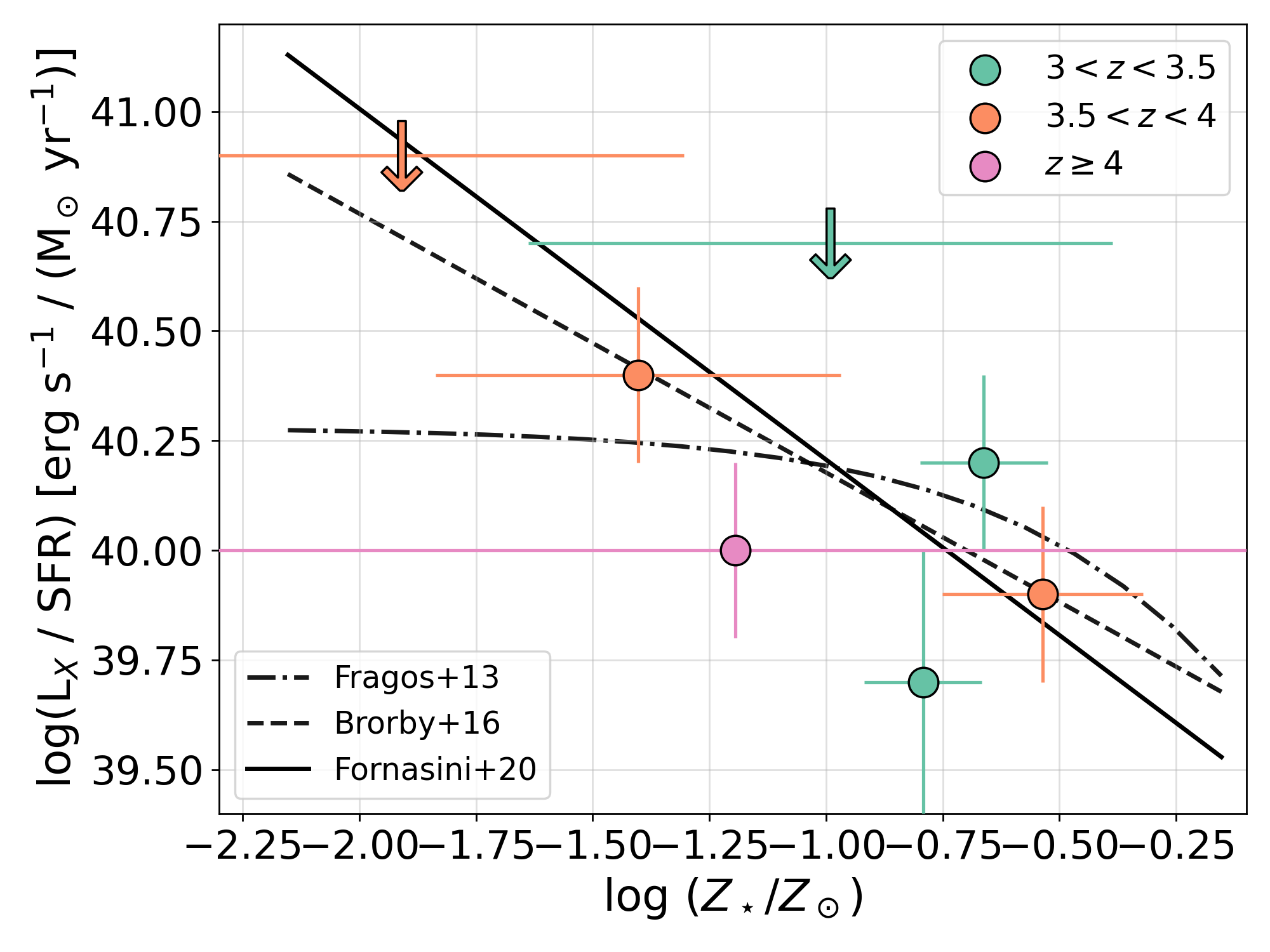}
    \caption{Distribution of stacked $L_X$/SFR measurements vs. stellar metallicity in SFR bins, where different colours represent the measurements in different redshift bins considered in this study (see Tables \ref{tab:subset_props} and \ref{tab:subsets_xray}). Also shown for comparison are theoretical predictions from \citealt{fra13a} (dot-dashed line), and empirical relations in both the local (dashed black line, \citealt{bro16}) and $z\sim1-2$ Universe (solid black line, \citealt{for20}). We find that our measurements are consistent within errors with both theoretical as well as empirical relations between $L_X$/SFR and metallicity. Our upper limits at the lowest metallicities support the expectations from models and the latest empirical fits from \citet{leh21} (not shown) that the relation deviates from a simple power-law and flattens out, similar to \citet{fra13a}.} 
    \label{fig:lx_sfr_metallicity}
\end{figure}

We begin our study of the relationship between $L_X$/SFR and stellar metallicity ($Z_\star$) measured following the methods described in \S \ref{sec:metallicity} by comparing our metallicity measurements with various theoretical and empirical models in the literature shown in Figure \ref{fig:lx_sfr_metallicity}. The coloured points represent $L_X$/SFR and $Z_\star$ measurements in the different redshift bins. The dot-dashed line shows the relationship between the two quantities obtained by \citet{fra13a} using HMXB modelling, the black dashed line shows the empirically derived relation from measurements in the local Universe \citep{bro16}, and the solid black line shows the empirical relation derived for $z\sim0-2$ galaxies \citep{for20}. 

We find that our metallicities and $L_X$/SFR measurements at $z>3$ are consistent within the errors with both theoretical and empirical models in the literature. The difference between the \citet{fra13a} model and the empirical power-law fits derived from low redshift galaxies is most significant at low metallicities, where the \citet{fra13a} model predicts a levelling-off of $L_X$/SFR. \citet{leh21} recently attempted to model the global metallicity dependence of HMXB emission by combining available observations at low redshifts, and found that qualitatively, their best-fit model follows a power-law at high metallicities but predicts a deviation from that power-law and a levelling-off of $L_X$/SFR at low metallicities (similar to the \citealt{fra13a} model). At the lowest metallicities in particular, we can only set upper limits on $L_X$/SFR, which make our measurements consistent with a higher order polynomial fit resulting in lower $L_X$/SFR values at low metallicities \citep[e.g.][]{leh21} as well as those predicted by power laws \citep[e.g.][]{dou15, bro16, pon20, for19, for20}. 

To parametrise the metallicity dependence of $L_X$/SFR, we consider a power-law relationship between $L_X$/SFR and stellar metallicity ($Z_\star$) for simplicity, which is written in terms of solar metallicity ($Z_\odot$) as:
\begin{equation}
    \log(L_X/\textrm{SFR}) = a + b\log(Z_\star/Z_\odot)
    \label{eq:metallicity}
\end{equation}

\begin{figure*}
    \centering
    \includegraphics[width=0.7\textwidth]{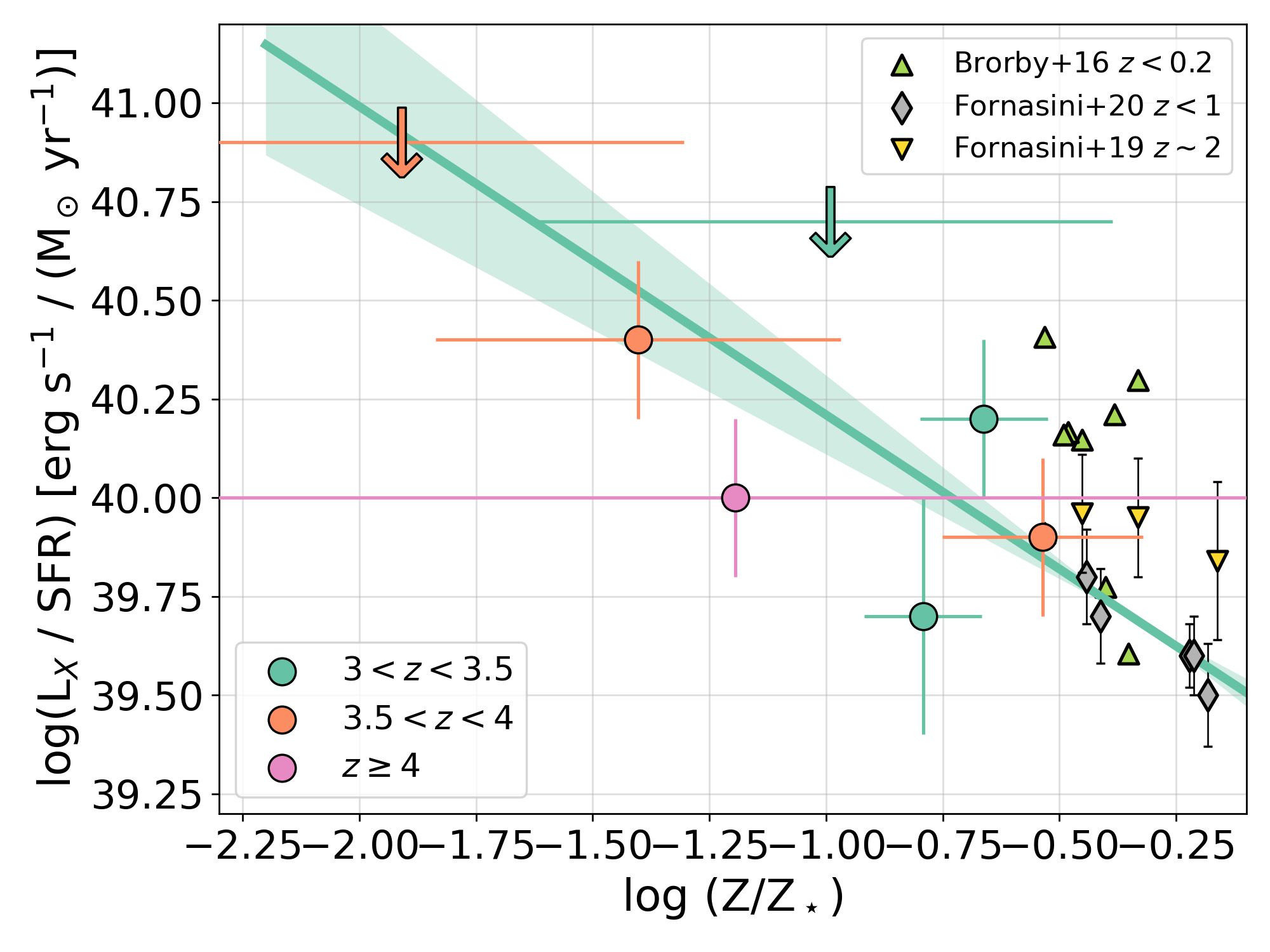}
    \caption{Best-fit relation between $L_X$/SFR and stellar metallicity, $Z_\star$ derived by combining our $z>3$ measurements with those from the literature at lower redshifts. Literature measurements include \citet{bro16} ($z<0.2$), \citet{for20} ($z<1$) and \citet{for19} ($z\sim2$). The best-fit derived is shown as a solid green line, with the shaded region marking the $2\sigma$ dispersion. We find that the best-fit coefficients obtained by adding our $z>3$ measurements are consistent with those obtained using only lower redshift measurements. The addition of our new measurements at $z>3$ probing much lower metallicities are crucial to constrain the metallicity dependence of $L_X$/SFR over several orders of magnitude. The best-fit coefficients and their comparison with other measurements in the literature can be found in the text.}
    \label{fig:metallicity_dependence}
\end{figure*}

To obtain reliable fits over several orders of magnitude of $Z_\star$, we also include measurements from lower redshifts, such as \citet{bro16} at $z<0.2$ and measurements from the metallicity bins from both \citet{for20} at $z<1$ and \citet{for19} at $z\sim2$. The distribution of $L_X$/SFR and $Z_\star$ of our $z>3$ sample along with measurements from the literature at lower redshifts is shown in Figure \ref{fig:metallicity_dependence}. We note here that the low redshift literature studies have utilised gas-phase metallicity measurements (12 + $\log(\textrm{O/H})$) derived from rest-frame optical emission lines. Therefore, we convert these gas-phase (O/H) ratios to metal mass fraction $Z$, using $Z=\textrm{(O/H)}*(\textrm{H}_\textrm{frac}/\textrm{O}_\textrm{frac})$, where $\textrm{H}_\textrm{frac}$ is the mass fraction of Hydrogen and $\textrm{O}_\textrm{frac}$ is the mass fraction of Oxygen. \citet{sax20b} found that setting the mass fraction of H to be 75\%, which is consistent with the composition of the sun, a mass fraction of $\approx40\%$ for O contributing to the observed metallicity gives consistent results when recovering the solar metallicity values for both (O/H) and $Z_\star$. We note here, however, that the absolute normalisations of the solar values of both $Z_\star$ and (O/H) are uncertain, and the conversion we derive serves as a purely empirical conversion.

We use non-linear least squares to obtain a best-fit relation between $L_X$/SFR and $Z_\star$ (Equation (\ref{eq:metallicity}) for the combined sample containing our $z>3$ measurements and the lower redshift measurements from the literature. We give more weight to our $z>3$ measurements in the fitting, as these represent the crucial low metallicity points. We find that using an orthogonal distance regression to find the best-fit function leads to highly consistent results. The coefficients of the best-fit power law are: $a=39.43 \pm 0.05$ and $b=-0.78 \pm 0.15$. The best-fit power law is shown in Figure \ref{fig:metallicity_dependence} (solid green line), along with the dispersion in this best-fit (shaded region).

As seen in Figure \ref{fig:metallicity_dependence}, our $z>3$ low metallicity measurements are crucial to constrain the $L_X$/SFR relation owing to the relatively large scatter within the low redshift measurements from the literature. We find that the best-fit power law we obtain is consistent with studies of lower redshift galaxies. For example, \citet{dou15} found a stronger evolution of $L_X$/SFR at lower metallicities for galaxies in the local Universe, with the metallicity scaling coefficient $b=-1.01$ with a dispersion of $\sim0.5$ dex. \citet{bro16} obtained $b=-0.64$ for their sample of star-forming galaxies at $z<0.12$, and \citet{pon20} reported $b\approx-0.95$ for their sample of compact dwarf galaxies in the local Universe. \citet{for20} reported $b\approx -0.80$ obtained by for star-forming galaxies at $z<1$, which is highly consistent with our measurements. 

This metallicity dependence of HMXB emission has been postulated to drive the redshift evolution of $L_X$/SFR \citep[see][for example]{bas13a}. The stellar (and gas-phase) metallicities are expected to decrease with increasing redshifts, which would in turn lead to stronger X-ray luminosities produced by HMXB populations per SFR. In the following section we compare our X-ray observations at $z>3$ with semi-analytical models to test whether the metallicity dependence we derive can reproduce the redshift evolution of $L_X$/SFR that we see in our $z>3$ measurements.

\subsection{Is redshift evolution of HMXB emission purely driven by metallicities?}
In this section, we employ stellar metallicities from the Galaxy Evolution and Assembly (GAEA) semi-analytical model, which tracks the formation, evolution and chemical enrichment of galaxies across cosmic time \citep{del07} to test whether a redshift evolution in metallicities can explain the redshift evolution of $L_X$/SFR as well. Using VANDELS spectroscopic data, \citet{cal21} reported close agreement between the observed slope of the mass-metallicity relation with predictions from GAEA models, with a consistent offset of 0.27 dex. This previously observed close agreement between VANDELS observations and GAEA has motivated the choice of using GAEA models to compare against observations in this study.

To build a comparison sample, we have assembled star-forming galaxies from GAEA that match the stellar mass and SFR distribution of our VANDELS galaxies. We divide GAEA galaxies in the same bins of SFR and redshift that were used for observations. In each SFR and redshift bin, we calculate the expected $L_X$/SFR based on the best-fit $L_X$-SFR-$Z_\star$ relation obtained using Equation (\ref{eq:metallicity}), taking the $1\sigma$ standard deviation of the distribution of $Z_\star$ values of GAEA galaxies to calculate the dispersion. In line with the findings of \citet{cal21}, we apply the 0.27 dex correction to GAEA stellar metallicities to enable accurate comparison with observations.
\begin{figure*}
    \centering
    \includegraphics[width=\textwidth]{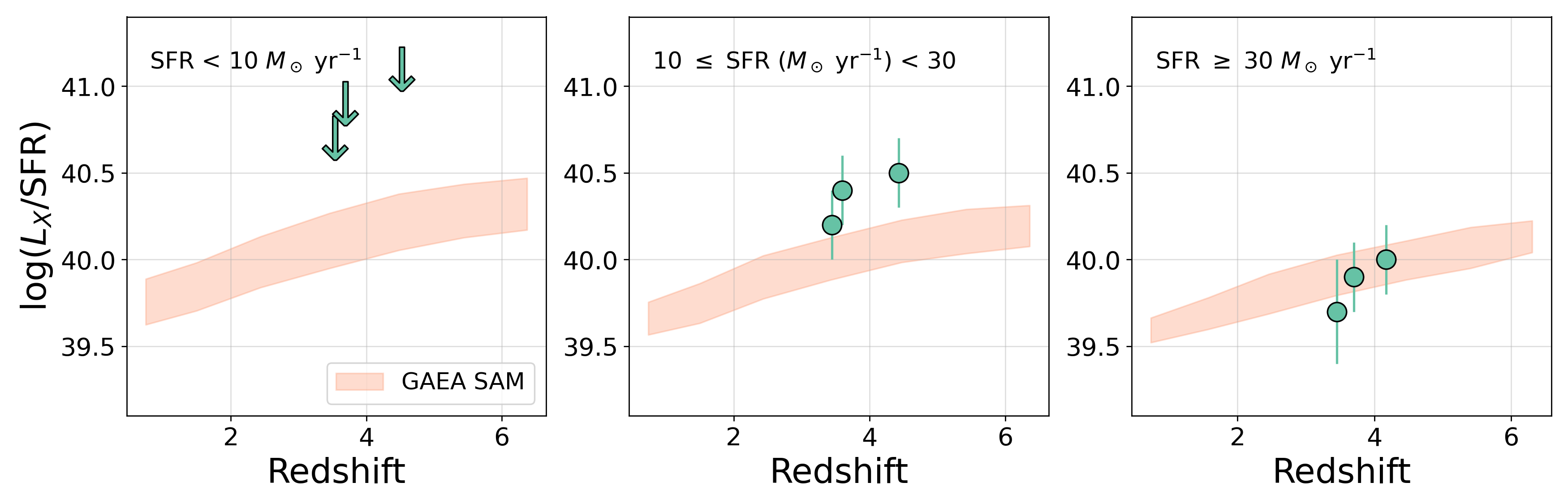}
    \caption{Observed redshift evolution of $L_X$/SFR at $z>3$ compared to the expected X-ray emission from the stellar metallicity distribution of galaxies in the GAEA \citep{del07} semi-analytical model (orange). We find that the metallicity evolution of galaxies can fully explain the observed redshift evolution of $L_X$/SFR for galaxies with SFR $>30$ \sfr, but the observed X-ray emission from HMXB populations in galaxies with SFR $<30$ \sfr\ is higher than expectations. This suggests that specifically for the low SFR, low metallicity galaxies, only the evolution of metallicities may not be enough to explain the redshift evolution of $L_X$/SFR.}
    \label{fig:gaea_comparison}
\end{figure*}

In Figure \ref{fig:gaea_comparison} we show our observed $L_X$/SFR measurements as a function of redshift in the three SFR bins, along with the range of expected $L_X$/SFR calculated using the GAEA stellar metallicities (orange shaded region). We find that there is good agreement between the observed redshift evolution of $L_X$/SFR for galaxies with SFR $>30$ \sfr\ and that expected from the evolving metallicities of GAEA model galaxies, suggesting that the global distribution of galaxy stellar metallicities across redshifts can explain the redshift evolution of X-ray emission from HMXBs \citep[see also][]{bas13a, leh16, for20}. However, for galaxies with SFR $<30$ \sfr\ we find that the $L_X$/SFR expected purely from the evolution of stellar metallicities is less than what the observations suggest. The $L_X$/SFR measured in the redshift bin $3.0 \leq z < 3.5$ for galaxies with SFR $=10-30$\,\sfr\ is consistent within the errors with the model predictions, but measurements at $z>3.5$ are $\sim0.2$ dex higher than the expected X-ray emission from evolving metallicities at high redshifts. X-ray observations in the SFR $<10$\,\sfr\ bin are only upper limits, making it impossible to compare with models for specifically the lowest SFR systems. However, following the trends of an increasing normalisation of $L_X$ per SFR with decreasing SFRs at the same redshift suggest that the model predictions may under-predict observations even in the lowest SFR bin. Therefore, the redshift evolution of emission from HMXBs purely due to evolving metallicities may not be enough specifically for the lower mass, lower stellar metallicity galaxy populations at the highest redshifts.

\subsection{A possible redshift evolution of the $\mathbf{L_X}$/SFR-$\mathbf{Z_\star}$ relation?}
This discrepancy between model predictions and observations suggests that other redshift-dependent galaxy properties may play a role in determining the evolution of HMXBs emission specifically at low metallicities. A possible physical explanation for the higher observed $L_X$/SFR values could be the sensitivity of $L_X$/SFR from HMXB populations to stellar ages \citep[see][for example]{rap05}. For the same stellar metallicity, \citet{sch19} showed, using the \citet{fra13a, fra13b} HMXB models, that the $L_X$/SFR can vary over up to three orders of magnitude from stellar ages of 0.01 Gyr to 10 Gyr. Furthermore, the dynamic range of $L_X$/SFR as a function of stellar age is larger at lower metallicities, which implies that $L_X$/SFR is more sensitive to changes in the stellar ages at lower stellar metallicities than at higher metallicities (see Figure 2 of \citealt{sch19} for example). 

An additional dependence of $L_X$/SFR on the stellar ages may be investigated by looking for any possible redshift dependence of the $L_X$-SFR-$Z_\star$ relation by inserting a redshift term to Equation (\ref{eq:metallicity}), such that:
\begin{equation}
    \log(L_X/\textrm{SFR}) = a + b\log(Z_\star/Z_\odot) + c\log(1+z)
    \label{eq:metallicity_redshift}
\end{equation}

We once again include measurements from \citet{bro16}, \citet{for19} and \citet{for20} for $z<2$ spectroscopically confirmed galaxies with metallicity measurements and use non-linear least squares to find the best-fit. We obtain the best-fit coefficients $a=39.42 \pm 0.03$, $b=-0.76 \pm 0.15$ and $c=0.23 \pm 0.19$, where the positive value of the redshift evolution coefficient $c$ suggests a possible redshift evolution in the $L_X$-SFR-$Z_\star$ relation. 

This redshift dependence could be explained by galaxies in our subsamples tracing different average star-formation histories (SFHs). In such a scenario, a stronger evolution of the X-ray luminosity per star-formation rate at high redshifts could be a result of overall lower stellar ages at high redshifts or `burstier' star-formation at early epochs, owing to the relatively short lifetimes ($5-100$\,Myr) of HMXBs. Different SFHs could also result in differences in the scaling of X-ray luminosity with SFR \citep[e.g.][]{kov20}, hints of which we see from our earlier analysis in \S\ref{sec:Redshift evolution of L_X/SFR}. Further, there is some suggestion in the literature that the slope of the $L_X$/SFR--$Z_\star$ relation changes with redshift; \citet{for20} found slightly different metallicity dependencies for their $z\sim0.2$ and $z\sim0.3$ samples, but these differences were not found to be statistically significant. Unfortunately, our current $z>3$ X-ray measurements lack the accuracy to enable simultaneously fitting the dependence of $L_X$ with SFRs, redshifts and metallicities, which may shed some light on the redshift and metallicity dependence of the scaling relationship between $L_X$ and SFR themselves, but this may be possible for HMXB populations at lower redshifts.

Nonetheless, we find that including the redshift evolution of $L_X$-SFR-$Z_\star$ relation results in a $0.2$ dex increase in the predicted $L_X$/SFR of HMXB populations at high redshifts, better explaining the observations for galaxies with SFR $=10-30$ \sfr\ when compared to the GAEA model predictions. Inclusion of this redshift evolution also provides consistent predictions for SFR $>30$ \sfr\ galaxies, showing that the redshift dependence of the $L_X$-SFR-$Z_\star$ relation is likely related to an additional dependence on SFHs or stellar ages. We note, however, that the significance of the redshift evolution of the $L_X$-SFR-$Z_\star$ relation is very low (marginally over $1\sigma$), but may represent a significant effect at very high redshifts where it is currently impossible to constrain the outputs of HMXB populations.

A drawback of deriving a redshift dependence from samples compiled from various data sets across redshifts is the underlying difference in key galaxy properties such as metallicities, star-formation rates and stellar masses, which play important roles in affecting the normalisation of $L_X$ per SFR. For example, the metallicities probed by our galaxy samples are systematically lower than those of samples at lower redshifts. Additionally, the SFR distribution of VANDELS galaxies at $z>3$ extends to larger values than that for the low redshift samples. Clearly, to reliably probe any redshift dependence on HMXB scaling relations, a key requirement is that the samples being probed (i) span a large redshift range, (ii) are uniformly selected in star-formation rates, and (iii) have comparable metallicities. However, this study demonstrates the best constraints that could be obtained at $z>3$ with current limitations on deep spectroscopic as well as X-ray data, and deeper spectroscopic surveys in the future combined with X-ray data from the next-generation telescopes will be crucial to enhance these constraints.

\section{HMXB emission from strong Lyman-alpha emitters}
\label{sec:XRB emission from strong Lyalpha emitters}
To contextualise the possible impact of HMXBs within galaxies at $z>6$ that are often considered the drivers of cosmic reionisation \citep{rob13}, in this section we explore the HMXB emission from a subsample of strong \lya\ emitting galaxies (LAEs) in our sample. LAEs are often considered valuable analogues of galaxies in the reionisation era given their low metallicities, compact morphologies and intense line emission \citep{fle19}. LAEs at intermediate redshifts are also found to have a higher production of ionising photons compared to the general Lyman break galaxy sample \citep[see][for example]{nak18}. The \lya\ line enters the observed wavelength range of VANDELS spectra at $z>3$ and here we investigate any possible excess X-ray emission from such galaxies, which could lead to an increase in both the production and escape of ionising photons from equivalent sources at $z>6$.

As mentioned previously, the LAEs in this analysis are selected from our global VANDELS galaxy sample at $z>3$ within the CDFS 7Ms X-ray footprint. The \lya\ emission line fluxes and equivalent widths are measured using the \textsc{python} package \textsc{mpdaf} (Muse Python Data Analysis Framework, \citealt{mpdaf}). The line flux and associated errors are estimated by fitting a single Gaussian to the emission line. Since the observed \lya\ emission is usually offset from its systemic redshift due to resonant scattering \citep[e.g.][]{ver18}, we allow for the peak of the best-fit Gaussian to be redshifted from the expected position from the systemic redshift by up to $\approx 2.4$\,\AA, the equivalent of $600$\,km\,s$^{-1}$. The width of the Gaussian may extend to 0.5\,\AA\ bluewards and 1.5\,\AA\ redwards from this range, given the generally asymmetric \lya\ line profiles observed. We estimate the continuum using a fifth-order polynomial in the wavelength range 1221.82\,\AA\ $\leq \lambda \leq 1236.1$\,\AA, which lies just redwards of the range used for line fitting, to avoid contamination. EW$_0$ is then calculated by dividing the line flux with the best-fit continuum, propagating their respective errors.
\begin{figure}
    \centering
    \includegraphics[scale=0.42]{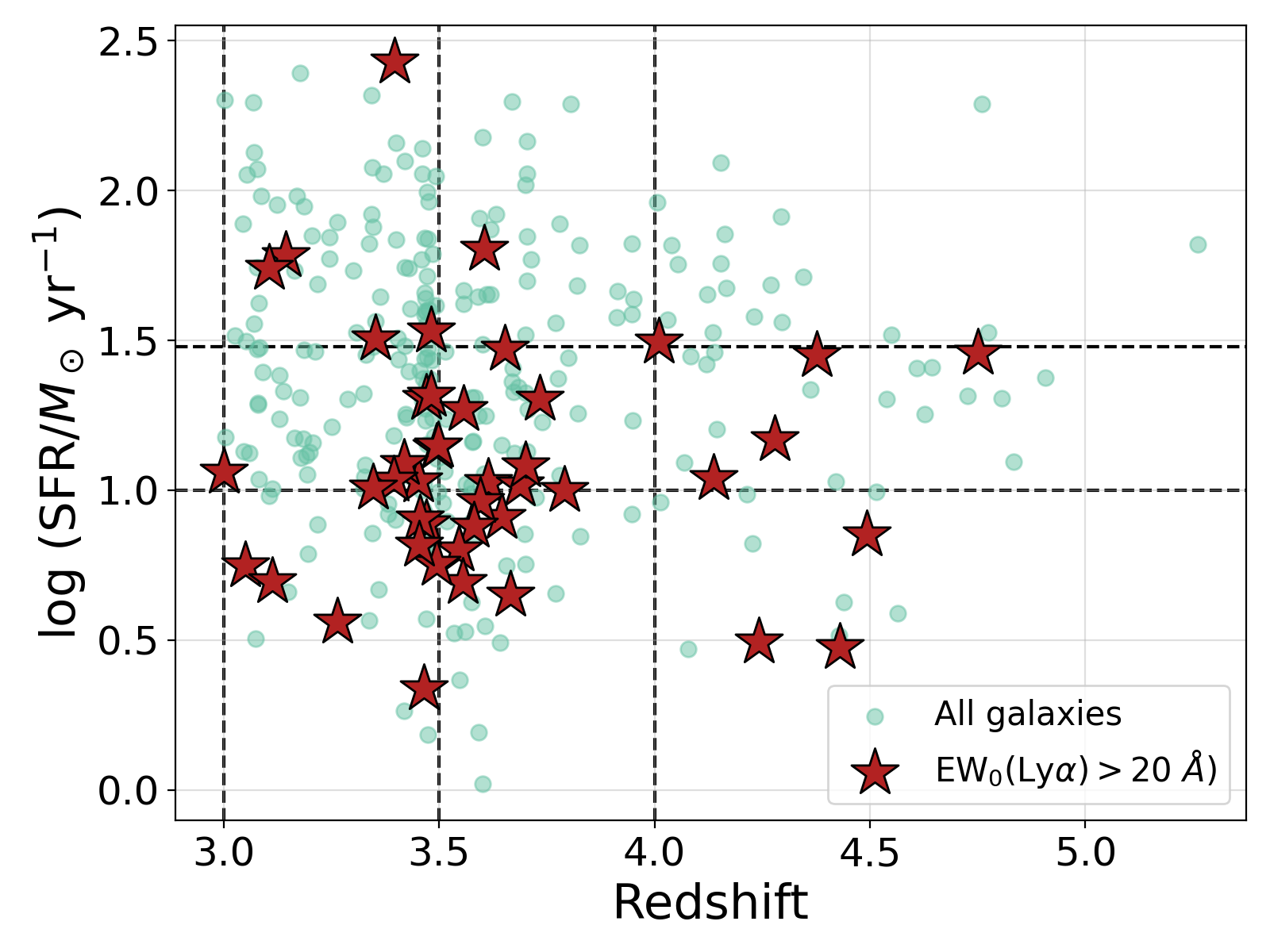}
    \caption{Distribution of redshifts and star-formation rates of strong LAEs compared to the parent population studied in previous sections. LAEs tend to occupy the region of the parameter space with low SFRs across redshifts. We divide the LAEs into redshift bins but not SFR bins as was done for the parent sample. We note the patchy sampling of the parameter space specifically in the $3<z\leq3.5$ bin.}
    \label{fig:lae_sample}
\end{figure}
\begin{table*}
    \centering
    \caption{Median redshifts, SFRs, stellar masses, specific SFRs and metallicities along with stacked X-ray luminosities and X-ray luminosities per SFR of LAEs (EW$_0 > 20$\,\AA) in redshift bins.}
    \begin{tabular}{l c c c c c c c c}
    \hline 
    Subset & N & Redshift & SFR (\sfr) & $\log(M_\star/M_\odot)$ & $\log$(sSFR/yr$^{-1}$) & $\log(Z_\star/Z_\odot)$ & $L_X$ ($\times 10^{41}$ \lum) & $\log$($L_X$/SFR) \\
    \hline \hline
   
    $3 \le z < 3.5$ & 22 & 3.44 & 11.2 & 9.0 & $-7.99$ & $-0.91 \pm 0.38$ & $3.4 \pm 1.5$ & $39.6 \pm 0.6$    \\
    
    $3.5 \le z < 4$ & 14 & 3.63 & 10.2 & 8.9 & $-7.93$ & $-1.52 \pm 1.02$ & $5.1 \pm 1.6$ & $40.2 \pm 0.5$ \\
    
    $z > 4$ & 8 & 4.33 & 12.7 & 9.0 & $-7.91$ & $-$ & $6.2 \pm 1.8$ & $40.4 \pm 0.5$ \\
    \hline
    \end{tabular}
    \label{tab:lae_subsets}
\end{table*}
 
We define our strong \lya\ emitting (LAE) sample using a rest-frame equivalent width cut of EW$_0 > 20$\,\AA, with errors on $\Delta \mathrm{EW}_0 <$ 0.7\,\AA. All sources that qualified as strong LAEs were visually inspected and confirmed. This selection results in 44 galaxies that can be classified as {\it bona fide} LAEs within the CDFS footprint. To place the physical properties of the LAE sample into context, we show the distribution of the redshifts and star-formation rates of LAEs compared with the distribution of our parent sample in Figure~\ref{fig:lae_sample}. We note that LAEs tend to have lower SFRs than the parent VANDELS sample.

The LAEs are divided in the same redshift bins as the parent sample: $z = 3-3.5$, $z=3.5-4$ and $z>4$. The X-ray photometry and stacking is performed in the same was as described for the parent sample in \S \ref{sec:X-ray photometry}. Table \ref{tab:lae_subsets} shows the median physical properties and the stacked X-ray measurements for the LAE subsamples. Since there are fewer LAEs overall, both stellar metallicities as well as stacked $L_X$ measurements have larger uncertainties. However, we find that the physical properties of LAEs across redshifts are highly comparable with the low to medium SFR galaxies in the parent sample, which in turn indicates that low-mass, lower metallicity galaxies often show strong \lya\ emission in their spectra.

\begin{figure}
    \centering
    \includegraphics[scale=0.42]{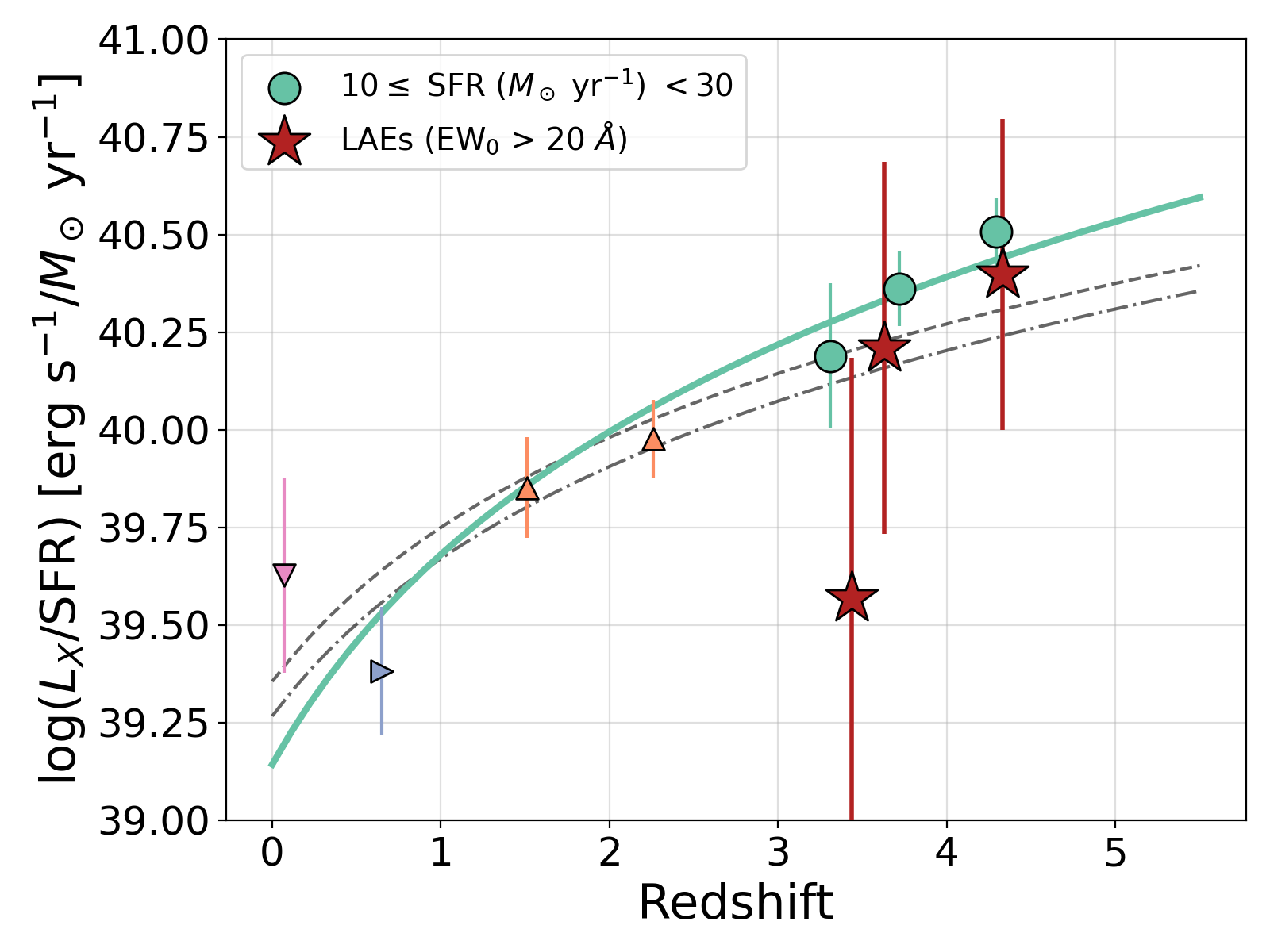}
    \caption{Comparison of $L_X$/SFR of \lya\ emitters (LAEs; EW$_0$(\lya) $>20$\,\AA) with stacked measurements in our medium SFR bins. We also show other measurements from the literature for comparison, where the symbols are the same as in Figure \ref{fig:lx_sfr_z_evol}. We do not find any statistically significant difference between $L_X$/SFR values of LAEs and those of galaxies with SFRs in the range $10-30$ \sfr for $z>$3.5.}
    \label{fig:LAEs}
\end{figure}

In Figure \ref{fig:LAEs} we show the measured $L_X$/SFR in the same redshift bins compared to our stacked measurements from the medium SFR ($10 <$\, SFR (\sfr)\,$< 30$) bins, along with other lower redshift measurements from the literature (as in Figure \ref{fig:lx_sfr_z_evol}). We find that LAEs at $z>3.5$ show highly comparable $L_X$/SFR values to those of the parent sample in the intermediate SFR bin, albeit with larger error bars owing to the lesser number of LAEs per redshift bin. We note here that the $3 < z < 3.5$ measurement is almost 0.5 dex lower than the corresponding measurement from that measured from the parent sample, and we attribute this to the relatively incomplete and sparse sampling of LAEs in this particular redshift bin, which leads to increased uncertainties and potential biases in the measurement.

Overall, we do not find any surprising results from the LAE sample. Moreover, we do not find any strong correlations between \lya\ strengths and X-ray fluxes, but we are also severely limited at these redshifts by the survey depth for individual X-ray detections. We conclude that galaxy properties such as SFRs, metallicities and specific SFRs are more likely to dominate the observed $L_X$/SFR scaling seen in galaxy populations, with strong \lya\ emission being a consequence of these galaxy properties \citep[see][for example]{sha03}. Rest-frame UV spectroscopy of lower redshift LAE analogues with existing X-ray data may hold the key to explore direct connections between \lya\ line strengths and HMXB emission.

\section{Contribution to the cosmic X-ray background}
\label{sec:Discussion}

Having constrained the redshift as well as metallicity dependence of the X-ray output of HMXB populations in $z>3$ galaxies, in this section we turn our attention to what these constraints mean for the global contribution of HMXBs at high redshifts to the cosmic X-ray background.
\begin{figure*}
    \centering
    \includegraphics[width=0.75\textwidth]{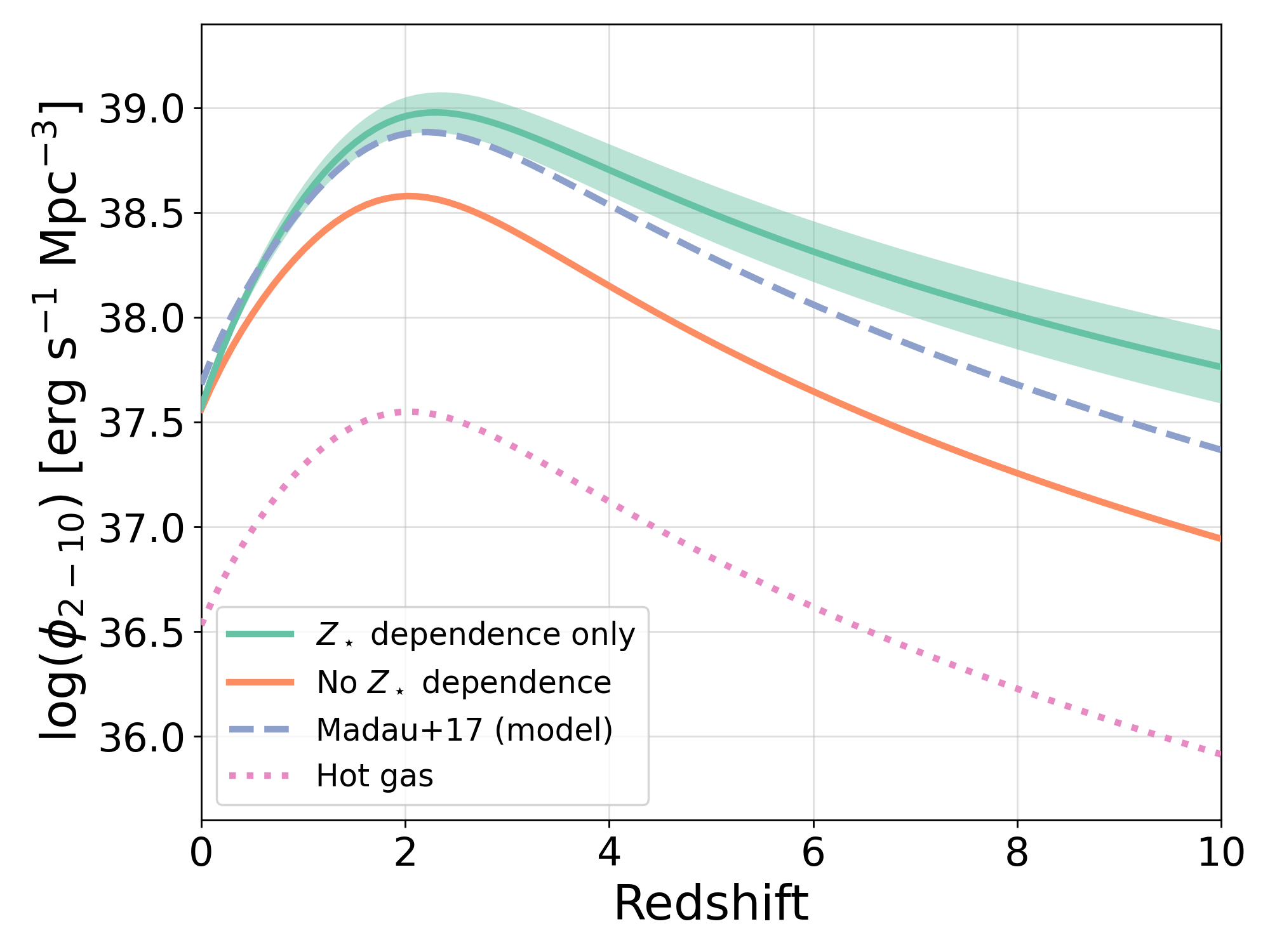}
    \caption{Integrated X-ray luminosity density in the $2-10$ keV energy range from HMXBs as a function of redshift. The emissivity is calculated assuming the cosmic SFR density evolution from \citet{mad14}. The solid green line shows the luminosity density calculated from our new constraints on $L_X$/SFR and its dependence on stellar metallicity at $z>3$, with the shaded region marking the $2\sigma$ dispersion. The solid orange line shows the redshift evolution of the luminosity density expected when no metallicity dependence of HMXB emission is assumed, normalised at $z=0$. The dotted pink line shows the contribution to the X-ray background expected from hot gas, following \citet{leh16}, which is considerably lower than the HMXB contribution. For comparison, we show the luminosity density from the \citet{mad17} models, and find that our new constraints indicate a $\gtrsim0.25$ dex increase in the HMXB contribution to the cosmic X-ray background at $z>6$.}
    \label{fig:xrb_emissivity}
\end{figure*}

An important quantity that traces the global contribution from a class of sources to the X-ray background is the X-ray luminosity density, $\phi$. Constraining the cosmic X-ray background from HMXBs can help study their role in the X-ray heating of the early Universe \citep[e.g.][]{mad17}. Studies have shown that the inclusion of sources of X-ray heating in reionisation calculations and simulations leads to a more extended and uniform reionisation history of the Universe \citep[e.g.][]{mes13}, higher IGM temperature fluctuations in the early Universe resulting in increased power of the 21cm signal at larger scales \citep[e.g.][]{war09, pac14}, and a possible preheating of the IGM in the Universe, which would precede the epoch of reionisation \citep[e.g.][]{mad17, mei17, eid18}. 

All of these effects can have an impact on the patchiness as well as the timescale of reionisation, which remain outstanding questions and can in principle be tested through 21cm experiments using the Square Kilometre Array \citep[see][]{mel13}. Therefore, constraining the X-ray luminosity densities at the highest redshifts has bearing on the reionisation history of the Universe. Although a full analysis of the contribution from HMXBs to the reionisation budget in light of our new constraints is beyond the scope of the current paper, we can derive a qualitative understanding of the implications of HMXBs on reionisation.

Using the metallicity dependence of $L_X$/SFR at $z \geq3$, we estimate the X-ray luminosity density in the $2-10$ keV energy band by combining the best-fit metallicity dependence of $L_X$/SFR and the evolution in the observed stellar metallicities at $z>3$ for HMXB populations seen in this study, with the redshift evolution of the cosmic SFR density, $\psi(z)$, from \citet{mad14} in the following way:
\begin{equation}
    \log(\phi_{2-10}) = \log[\psi(z)] + \log \left[\frac{L_X}{\rm{SFR}}(Z_\star)\right]
\end{equation}
where $L_X$/SFR is evaluated following the best-fit metallicity relations from Equation (\ref{eq:metallicity}). For simplicity, we do not include the redshift evolution term from Equation (\ref{eq:metallicity_redshift}) as the redshift evolution coefficient was not found to be significant.

The solid green line in Figure \ref{fig:xrb_emissivity} shows the redshift evolution of X-ray luminosity density obtained using new constraints at $z>3$, with the shaded region representing the $2\sigma$ dispersion. The solid orange line shows the expected evolution in the X-ray luminosity density when we ignore any $Z_\star$ dependence of $L_X$/SFR, whereas the blue dashed line shows the predictions from the \citet{mad17} model largely based on the metallicity dependence modelled by \citet{fra13a}. As shown by \citet{leh16} and \citet{mad17}, HMXBs become the dominant contributors to the X-ray background at $z>5$, since the contribution from AGN declines significantly \citep{air15}. The contribution of hot gas to the background at high redshifts (dotted pink line), calculated using the prescription laid out by \citet{leh16} based on the work of \citet{min12}, remains $\sim1.5$ dex lower than that of HMXBs. The low-mass X-ray binary population is also not expected to be a dominant contributor beyond $z>5$, as HMXBs outshine lower-mass ones already at $z>2$ \citep{mad17}.

Our new constraints on the metallicity dependence and the redshift evolution of the HMXB populations at $z>3$ suggest that their net contribution to the cosmic X-ray background at $z>6$ may be marginally higher than previously modelled. We find a $\sim0.25$ dex increase in the X-ray luminosity density due to HMXBs at $z=6$, and the functional form of the redshift evolution of the luminosity density suggests a rising increase at even higher redshifts compared to the predictions of \citet{mad17}. At $z\sim10$, for example, our new constraints suggest a $\sim0.4$ dex increase compared to \citet{mad17}. This has implications on the X-ray heating provided by the output of HMXBs at the epochs preceding reionisation, in the era of Population III stars for example, where the expected X-ray luminosities and their subsequent impact on the 21cm signatures may be higher \citep[see][for example]{mes13}.

In terms of the impact of X-ray photons from HMXBs on reionisation at galactic scales, \citet{mad17} argued that the contribution of HMXBs, from both soft and hard X-rays, remains negligible compared to the ionising output of low metallicity (e.g. $Z_\star = 1/20\,Z_\odot$) stars at $z>6$, which is $\sim2.5$ dex higher than that of HMXBs. Considering even the lowest possible limits on the escape fraction of ionising radiation from stars, the HMXB contribution remains highly fractional. The $\gtrsim 0.25$ dex increase in the global emissivity of HMXBs that we report will not significantly impact the ionising output and contribution to reionisation from such systems, with the emerging picture being that radiation from HMXBs in the early Universe is more likely to have played an important role in shaping the reionisation history of the Universe.

\section{Conclusions}
\label{sec:Conclusions}

In this study we have presented, for the first time, new constraints on the X-ray luminosity from the high-mass X-ray binary (HMXB) populations based on spectroscopically confirmed galaxies at $z>3$. The galaxies have been selected from the recently completed VANDELS spectroscopic survey in the \textit{Chandra} Deep Field South (CDFS), which also benefits from a total of 7 Ms of multi-epoch \textit{Chandra} and remains the deepest X-ray image available. Using this unique combination of data sets, we identify a total of 301 Lyman-break galaxies in the redshift range $3 < z < 5.5$  with reliable spectroscopic redshifts. We derive accurate physical properties such as stellar masses, dust attenuation and star-formation rates (SFR) using available exquisite broadband imaging from both ground-based and space-based observatories, ranging from UV to infrared wavelengths, which also help eliminate potential AGN in the sample.

Since the individual X-ray luminosities from HMXB populations in $z>3$ galaxies are expected to be faint, we must rely on stacking to boost signal. We do this by dividing our sample in redshift and star-formation rate bins and measuring stacked X-ray counts, fluxes and luminosities in the rest-frame energy range $2-10$ keV using aperture-corrected photometry. We also calculate the stacked X-ray luminosity per unit star-formation ($L_X$/SFR) in each bin, which is a tracer for the HMXB output of star-forming galaxies.

Thanks to deep VANDELS rest-frame UV spectra, we obtain reliable stellar metallicity measurements from stacks, pushing studies of the dependence of HMXB emission to lower stellar metallicities than has previously been possible. Additionally, we also explore X-ray emission specifically from strong \lya\ emitting galaxies (LAEs) selected from VANDELS, that are analogous to low-mass galaxies in the early Universe that were responsible for driving the bulk of reionisation. Below we summarise the key findings of this study:
\begin{itemize}
    \item We find $L_X = 2.7 - 6.6 \times 10^{41}$\,\ \lum\ for stacks of $z>3$ galaxies presented in this study. We find that $L_X$/SFR ranges between $10^{40.2-40.5}$\,\ \lum/\sfr at $z>3$ for galaxies with star-formation rates in the range $10-30$ \sfr, and between $10^{39.7-40.0}$\,\ \lum/\sfr\ for galaxies with SFR $>30$\,\ \sfr. The X-ray counts from galaxies with SFR $<10$\,\ \sfr\ at $z>3$ are too faint for reliable measurements, and therefore we place upper limits of $L_X$/SFR $<10^{40.7}$ \lum/\sfr.
    
    \item We see a strong redshift evolution in the $L_X$/SFR for galaxies with SFR in the range $10-30$ \sfr\ with $(1+z)^{1.78 \pm 0.09}$. This redshift evolution is compatible with the upper limits we obtain for galaxies with SFR $<10$ \sfr\ bin. The redshift evolution of $L_X$/SFR in the SFR $>30$ \sfr\ bin is weaker, with $(1+z)^{0.79 \pm 0.02}$. The overall redshift evolution of $L_X$/SFR seen across the full sample is  $(1+z)^{1.03\pm0.02}$.
    
    \item We further constrain the crucial stellar metallicity dependence of $L_X$/SFR at $z>3$, pushing to more than an order of magnitude lower stellar metallicity measurements ($Z_\star$) than previous studies at lower redshifts. We find a strong anti-correlation between $L_X$/SFR and $Z_\star$, which can be parameterised by a power law with index $-0.78 \pm 0.15$. 
    
    \item To test whether the metallicity dependence of $L_X$/SFR is enough to explain the observed redshift evolution of $L_X$/SFR, we employ stellar metallicity measurements of mock galaxies from a semi-analytical galaxy evolution model, which have been matched to our VANDELS sample in terms of physical properties. We find that the $Z_\star$ dependence of $L_X$/SFR alone is insufficient to explain the observed redshift evolution of $L_X$/SFR measured from galaxies with SFR $=10-30$ \sfr, but explains well the redshift evolution of galaxies with SFR $>30$ \sfr. Although not statistically required by the data, we find that the addition of a redshift dependence of the $L_X$-SFR-$Z_\star$ relation can better explain the redshift evolution of $L_X$/SFR for galaxies with lower SFRs.
    
    \item We use our new constraints on X-ray emission from HMXB populations in galaxies at $z>3$ to estimate their contribution to the cosmic X-ray background. We find that the metallicity dependence we observe predicts a $\gtrsim 0.25$ dex increase in the X-ray luminosity density of HMXBs at $z>6$ compared to previous predictions, which may have bearing on the cosmic reionisation history due to pre-heating of the Universe by X-ray photons emitted from HMXB populations in the early Universe.
\end{itemize}

Our new constraints on the X-ray luminosities of HMXB populations in star-forming galaxies at $z>3$ and their contribution to the global X-ray background represent a step forward towards characterising this population in the early Universe. Future X-ray missions such as the \textit{Advanced Telescope for High-ENergy Astrophysics (Athena)} and the \textit{Lynx X-ray Observatory}, in conjunction with galaxy spectra from \textit{JWST} have the potential to reveal some of the very first HMXB populations to have formed in the Universe.

\section*{Acknowledgements}
We thank the anonymous referee for constructive comments that helped improve the quality of this work. AS, RSE and PUF acknowledge funding from the European Research Council under the European Union Horizon 2020 research and innovation programme (grant agreement No 669253). AS would like to thank Marco Mignoli for useful discussions, and Fabrizio Fiore and Simonetta Puccetti for sharing the \textit{Chandra} data. BG acknowledges the support of Premiale MITIC 2017 and INAF PRIN Mainstream 2019. RA acknowledges support from ANID Fondecyt Regular 1202007. 

This work has made extensive use of \textsc{jupyter} and \textsc{ipython} \citep{ipython}, \textsc{astropy} \citep{astropy}, \textsc{matplotlib} \citep{plt}, \textit{pandas} \citep{pandas} and \textsc{topcat} \citep{topcat}. This work would not have been possible without the countless hours put in by members of the open-source developing community all around the world.

\section*{Data Availability}
The data underlying this article are part of VANDELS, which is a European Southern Observatory (ESO) Public Spectroscopic Survey. The data can be accessed using the VANDELS database at \url{http://vandels.inaf.it}, or through the ESO archives. The \textit{Chandra} data is publicly available from the \emph{Chandra} X-ray Centre \url{http://cxc.harvard.edu/cdo/cdfs.html} and is described in \citet{cdfs7ms}. The code used to perform the analysis in this paper can be shared upon reasonable request to the corresponding author.



\bibliographystyle{mnras}
\bibliography{XRB} 

\begin{thebibliography}{}
\makeatletter
\relax
\def\mn@urlcharsother{\let\do\@makeother \do\$\do\&\do\#\do\^\do\_\do\%\do\~}
\def\mn@doi{\begingroup\mn@urlcharsother \@ifnextchar [ {\mn@doi@}
  {\mn@doi@[]}}
\def\mn@doi@[#1]#2{\def\@tempa{#1}\ifx\@tempa\@empty \href
  {http://dx.doi.org/#2} {doi:#2}\else \href {http://dx.doi.org/#2} {#1}\fi
  \endgroup}
\def\mn@eprint#1#2{\mn@eprint@#1:#2::\@nil}
\def\mn@eprint@arXiv#1{\href {http://arxiv.org/abs/#1} {{\tt arXiv:#1}}}
\def\mn@eprint@dblp#1{\href {http://dblp.uni-trier.de/rec/bibtex/#1.xml}
  {dblp:#1}}
\def\mn@eprint@#1:#2:#3:#4\@nil{\def\@tempa {#1}\def\@tempb {#2}\def\@tempc
  {#3}\ifx \@tempc \@empty \let \@tempc \@tempb \let \@tempb \@tempa \fi \ifx
  \@tempb \@empty \def\@tempb {arXiv}\fi \@ifundefined
  {mn@eprint@\@tempb}{\@tempb:\@tempc}{\expandafter \expandafter \csname
  mn@eprint@\@tempb\endcsname \expandafter{\@tempc}}}

\bibitem[\protect\citeauthoryear{{Aird}, {Coil}, {Georgakakis}, {Nandra},
  {Barro}  \& {P{\'e}rez-Gonz{\'a}lez}}{{Aird} et~al.}{2015}]{air15}
{Aird} J.,  {Coil} A.~L.,  {Georgakakis} A.,  {Nandra} K.,  {Barro} G.,
  {P{\'e}rez-Gonz{\'a}lez} P.~G.,  2015, \mn@doi [\mnras]
  {10.1093/mnras/stv1062}, \href
  {https://ui.adsabs.harvard.edu/abs/2015MNRAS.451.1892A} {451, 1892}

\bibitem[\protect\citeauthoryear{{Aird}, {Coil}  \& {Georgakakis}}{{Aird}
  et~al.}{2017}]{air17}
{Aird} J.,  {Coil} A.~L.,   {Georgakakis} A.,  2017, \mn@doi [\mnras]
  {10.1093/mnras/stw2932}, \href
  {https://ui.adsabs.harvard.edu/abs/2017MNRAS.465.3390A} {465, 3390}

\bibitem[\protect\citeauthoryear{{Antoniou} \& {Zezas}}{{Antoniou} \&
  {Zezas}}{2016}]{ant16}
{Antoniou} V.,  {Zezas} A.,  2016, \mn@doi [\mnras] {10.1093/mnras/stw167},
  \href {https://ui.adsabs.harvard.edu/abs/2016MNRAS.459..528A} {459, 528}

\bibitem[\protect\citeauthoryear{{Asplund}, {Grevesse}, {Sauval}  \&
  {Scott}}{{Asplund} et~al.}{2009}]{asp09}
{Asplund} M.,  {Grevesse} N.,  {Sauval} A.~J.,   {Scott} P.,  2009, \mn@doi
  [\araa] {10.1146/annurev.astro.46.060407.145222}, \href
  {https://ui.adsabs.harvard.edu/abs/2009ARA&A..47..481A} {47, 481}

\bibitem[\protect\citeauthoryear{{Astropy Collaboration} et~al.,}{{Astropy
  Collaboration} et~al.}{2013}]{astropy}
{Astropy Collaboration} et~al., 2013, \mn@doi [\aap]
  {10.1051/0004-6361/201322068}, \href
  {http://adsabs.harvard.edu/abs/2013A%26A...558A..33A} {558, A33}

\bibitem[\protect\citeauthoryear{{Bacon}, {Piqueras}, {Conseil}, {Richard}  \&
  {Shepherd}}{{Bacon} et~al.}{2016}]{mpdaf}
{Bacon} R.,  {Piqueras} L.,  {Conseil} S.,  {Richard} J.,   {Shepherd} M.,
  2016, {MPDAF: MUSE Python Data Analysis Framework}, Astrophysics Source Code
  Library (\mn@eprint {ascl} {1611.003})

\bibitem[\protect\citeauthoryear{{Baldwin}, {Phillips}  \&
  {Terlevich}}{{Baldwin} et~al.}{1981}]{bal81}
{Baldwin} J.~A.,  {Phillips} M.~M.,   {Terlevich} R.,  1981, \mn@doi [\pasp]
  {10.1086/130766}, \href
  {https://ui.adsabs.harvard.edu/abs/1981PASP...93....5B} {93, 5}

\bibitem[\protect\citeauthoryear{{Basu-Zych} et~al.,}{{Basu-Zych}
  et~al.}{2013a}]{bas13a}
{Basu-Zych} A.~R.,  et~al., 2013a, \mn@doi [\apj] {10.1088/0004-637X/762/1/45},
  \href {https://ui.adsabs.harvard.edu/abs/2013ApJ...762...45B} {762, 45}

\bibitem[\protect\citeauthoryear{{Basu-Zych} et~al.,}{{Basu-Zych}
  et~al.}{2013b}]{bas13b}
{Basu-Zych} A.~R.,  et~al., 2013b, \mn@doi [\apj]
  {10.1088/0004-637X/774/2/152}, \href
  {https://ui.adsabs.harvard.edu/abs/2013ApJ...774..152B} {774, 152}

\bibitem[\protect\citeauthoryear{{Becker} \& {Bolton}}{{Becker} \&
  {Bolton}}{2013}]{bec13}
{Becker} G.~D.,  {Bolton} J.~S.,  2013, \mn@doi [\mnras]
  {10.1093/mnras/stt1610}, \href
  {https://ui.adsabs.harvard.edu/abs/2013MNRAS.436.1023B} {436, 1023}

\bibitem[\protect\citeauthoryear{{Bodaghee}, {Tomsick}, {Rodriguez}  \&
  {James}}{{Bodaghee} et~al.}{2012}]{bod12}
{Bodaghee} A.,  {Tomsick} J.~A.,  {Rodriguez} J.,   {James} J.~B.,  2012,
  \mn@doi [\apj] {10.1088/0004-637X/744/2/108}, \href
  {https://ui.adsabs.harvard.edu/abs/2012ApJ...744..108B} {744, 108}

\bibitem[\protect\citeauthoryear{Bradley et~al.,}{Bradley
  et~al.}{2019}]{photutils}
Bradley L.,  et~al., 2019, astropy/photutils: v0.6,
  \mn@doi{10.5281/zenodo.2533376}, \url
  {https://doi.org/10.5281/zenodo.2533376}

\bibitem[\protect\citeauthoryear{{Bromm} \& {Yoshida}}{{Bromm} \&
  {Yoshida}}{2011}]{bro11}
{Bromm} V.,  {Yoshida} N.,  2011, \mn@doi [\araa]
  {10.1146/annurev-astro-081710-102608}, \href
  {https://ui.adsabs.harvard.edu/abs/2011ARA&A..49..373B} {49, 373}

\bibitem[\protect\citeauthoryear{{Brorby}, {Kaaret}, {Prestwich}  \&
  {Mirabel}}{{Brorby} et~al.}{2016}]{bro16}
{Brorby} M.,  {Kaaret} P.,  {Prestwich} A.,   {Mirabel} I.~F.,  2016, \mn@doi
  [\mnras] {10.1093/mnras/stw284}, \href
  {https://ui.adsabs.harvard.edu/abs/2016MNRAS.457.4081B} {457, 4081}

\bibitem[\protect\citeauthoryear{{Bruzual} \& {Charlot}}{{Bruzual} \&
  {Charlot}}{2003}]{bc03}
{Bruzual} G.,  {Charlot} S.,  2003, \mn@doi [\mnras]
  {10.1046/j.1365-8711.2003.06897.x}, \href
  {https://ui.adsabs.harvard.edu/abs/2003MNRAS.344.1000B} {344, 1000}

\bibitem[\protect\citeauthoryear{{Bundy}, {Ellis}  \& {Conselice}}{{Bundy}
  et~al.}{2005}]{bun05}
{Bundy} K.,  {Ellis} R.~S.,   {Conselice} C.~J.,  2005, \mn@doi [\apj]
  {10.1086/429549}, \href
  {https://ui.adsabs.harvard.edu/abs/2005ApJ...625..621B} {625, 621}

\bibitem[\protect\citeauthoryear{{Calabr{\`o}} et~al.,}{{Calabr{\`o}}
  et~al.}{2021}]{cal21}
{Calabr{\`o}} A.,  et~al., 2021, \mn@doi [\aap] {10.1051/0004-6361/202039244},
  \href {https://ui.adsabs.harvard.edu/abs/2021A&A...646A..39C} {646, A39}

\bibitem[\protect\citeauthoryear{{Calzetti}, {Armus}, {Bohlin}, {Kinney},
  {Koornneef}  \& {Storchi-Bergmann}}{{Calzetti} et~al.}{2000}]{cal00}
{Calzetti} D.,  {Armus} L.,  {Bohlin} R.~C.,  {Kinney} A.~L.,  {Koornneef} J.,
   {Storchi-Bergmann} T.,  2000, \mn@doi [\apj] {10.1086/308692}, \href
  {https://ui.adsabs.harvard.edu/abs/2000ApJ...533..682C} {533, 682}

\bibitem[\protect\citeauthoryear{{Carnall}, {McLure}, {Dunlop}  \&
  {Dav{\'e}}}{{Carnall} et~al.}{2018}]{car18}
{Carnall} A.~C.,  {McLure} R.~J.,  {Dunlop} J.~S.,   {Dav{\'e}} R.,  2018,
  \mn@doi [\mnras] {10.1093/mnras/sty2169}, \href
  {https://ui.adsabs.harvard.edu/abs/2018MNRAS.480.4379C} {480, 4379}

\bibitem[\protect\citeauthoryear{{Chevallard} \& {Charlot}}{{Chevallard} \&
  {Charlot}}{2016}]{che16}
{Chevallard} J.,  {Charlot} S.,  2016, \mn@doi [\mnras]
  {10.1093/mnras/stw1756}, \href
  {https://ui.adsabs.harvard.edu/abs/2016MNRAS.462.1415C} {462, 1415}

\bibitem[\protect\citeauthoryear{{Circosta} et~al.,}{{Circosta}
  et~al.}{2019}]{cir19}
{Circosta} C.,  et~al., 2019, \mn@doi [\aap] {10.1051/0004-6361/201834426},
  \href {https://ui.adsabs.harvard.edu/abs/2019A&A...623A.172C} {623, A172}

\bibitem[\protect\citeauthoryear{{Colbert}, {Heckman}, {Ptak}, {Strickland}  \&
  {Weaver}}{{Colbert} et~al.}{2004}]{col04}
{Colbert} E. J.~M.,  {Heckman} T.~M.,  {Ptak} A.~F.,  {Strickland} D.~K.,
  {Weaver} K.~A.,  2004, \mn@doi [\apj] {10.1086/380899}, \href
  {https://ui.adsabs.harvard.edu/abs/2004ApJ...602..231C} {602, 231}

\bibitem[\protect\citeauthoryear{{Cullen} et~al.,}{{Cullen}
  et~al.}{2019}]{cul19}
{Cullen} F.,  et~al., 2019, \mn@doi [\mnras] {10.1093/mnras/stz1402}, \href
  {https://ui.adsabs.harvard.edu/abs/2019MNRAS.487.2038C} {487, 2038}

\bibitem[\protect\citeauthoryear{{De Lucia} \& {Blaizot}}{{De Lucia} \&
  {Blaizot}}{2007}]{del07}
{De Lucia} G.,  {Blaizot} J.,  2007, \mn@doi [\mnras]
  {10.1111/j.1365-2966.2006.11287.x}, \href
  {https://ui.adsabs.harvard.edu/abs/2007MNRAS.375....2D} {375, 2}

\bibitem[\protect\citeauthoryear{{Douna}, {Pellizza}, {Mirabel}  \&
  {Pedrosa}}{{Douna} et~al.}{2015}]{dou15}
{Douna} V.~M.,  {Pellizza} L.~J.,  {Mirabel} I.~F.,   {Pedrosa} S.~E.,  2015,
  \mn@doi [\aap] {10.1051/0004-6361/201525617}, \href
  {https://ui.adsabs.harvard.edu/abs/2015A&A...579A..44D} {579, A44}

\bibitem[\protect\citeauthoryear{{Eide}, {Graziani}, {Ciardi}, {Feng},
  {Kakiichi}  \& {Di Matteo}}{{Eide} et~al.}{2018}]{eid18}
{Eide} M.~B.,  {Graziani} L.,  {Ciardi} B.,  {Feng} Y.,  {Kakiichi} K.,   {Di
  Matteo} T.,  2018, \mn@doi [\mnras] {10.1093/mnras/sty272}, \href
  {https://ui.adsabs.harvard.edu/abs/2018MNRAS.476.1174E} {476, 1174}

\bibitem[\protect\citeauthoryear{{Eldridge}, {Stanway}, {Xiao}, {McClelland},
  {Taylor}, {Ng}, {Greis}  \& {Bray}}{{Eldridge} et~al.}{2017}]{eld17}
{Eldridge} J.~J.,  {Stanway} E.~R.,  {Xiao} L.,  {McClelland} L.~A.~S.,
  {Taylor} G.,  {Ng} M.,  {Greis} S.~M.~L.,   {Bray} J.~C.,  2017, \mn@doi
  [\pasa] {10.1017/pasa.2017.51}, \href
  {https://ui.adsabs.harvard.edu/abs/2017PASA...34...58E} {34, e058}

\bibitem[\protect\citeauthoryear{{Falc{\'o}n-Barroso},
  {S{\'a}nchez-Bl{\'a}zquez}, {Vazdekis}, {Ricciardelli}, {Cardiel}, {Cenarro},
  {Gorgas}  \& {Peletier}}{{Falc{\'o}n-Barroso} et~al.}{2011}]{fal11}
{Falc{\'o}n-Barroso} J.,  {S{\'a}nchez-Bl{\'a}zquez} P.,  {Vazdekis} A.,
  {Ricciardelli} E.,  {Cardiel} N.,  {Cenarro} A.~J.,  {Gorgas} J.,
  {Peletier} R.~F.,  2011, \mn@doi [\aap] {10.1051/0004-6361/201116842}, \href
  {https://ui.adsabs.harvard.edu/abs/2011A&A...532A..95F} {532, A95}

\bibitem[\protect\citeauthoryear{{Fan}, {Carilli}  \& {Keating}}{{Fan}
  et~al.}{2006}]{fan06}
{Fan} X.,  {Carilli} C.~L.,   {Keating} B.,  2006, \mn@doi [\araa]
  {10.1146/annurev.astro.44.051905.092514}, \href
  {https://ui.adsabs.harvard.edu/abs/2006ARA&A..44..415F} {44, 415}

\bibitem[\protect\citeauthoryear{{Fletcher}, {Tang}, {Robertson}, {Nakajima},
  {Ellis}, {Stark}  \& {Inoue}}{{Fletcher} et~al.}{2019}]{fle19}
{Fletcher} T.~J.,  {Tang} M.,  {Robertson} B.~E.,  {Nakajima} K.,  {Ellis}
  R.~S.,  {Stark} D.~P.,   {Inoue} A.,  2019, \mn@doi [\apj]
  {10.3847/1538-4357/ab2045}, \href
  {https://ui.adsabs.harvard.edu/abs/2019ApJ...878...87F} {878, 87}

\bibitem[\protect\citeauthoryear{{Fornasini} et~al.,}{{Fornasini}
  et~al.}{2019}]{for19}
{Fornasini} F.~M.,  et~al., 2019, \mn@doi [\apj] {10.3847/1538-4357/ab4653},
  \href {https://ui.adsabs.harvard.edu/abs/2019ApJ...885...65F} {885, 65}

\bibitem[\protect\citeauthoryear{{Fornasini}, {Civano}  \& {Suh}}{{Fornasini}
  et~al.}{2020}]{for20}
{Fornasini} F.~M.,  {Civano} F.,   {Suh} H.,  2020, \mn@doi [\mnras]
  {10.1093/mnras/staa1211}, \href
  {https://ui.adsabs.harvard.edu/abs/2020MNRAS.495..771F} {495, 771}

\bibitem[\protect\citeauthoryear{{Fragos} et~al.,}{{Fragos}
  et~al.}{2013a}]{fra13a}
{Fragos} T.,  et~al., 2013a, \mn@doi [\apj] {10.1088/0004-637X/764/1/41}, \href
  {https://ui.adsabs.harvard.edu/abs/2013ApJ...764...41F} {764, 41}

\bibitem[\protect\citeauthoryear{{Fragos}, {Lehmer}, {Naoz}, {Zezas}  \&
  {Basu-Zych}}{{Fragos} et~al.}{2013b}]{fra13b}
{Fragos} T.,  {Lehmer} B.~D.,  {Naoz} S.,  {Zezas} A.,   {Basu-Zych} A.,
  2013b, \mn@doi [\apjl] {10.1088/2041-8205/776/2/L31}, \href
  {https://ui.adsabs.harvard.edu/abs/2013ApJ...776L..31F} {776, L31}

\bibitem[\protect\citeauthoryear{{Garilli, B.} et~al.,}{{Garilli, B.}
  et~al.}{2021}]{gar21}
{Garilli, B.} et~al., 2021, \mn@doi [A\&A] {10.1051/0004-6361/202040059}, 647,
  A150

\bibitem[\protect\citeauthoryear{{Gehrels}}{{Gehrels}}{1986}]{geh86}
{Gehrels} N.,  1986, \mn@doi [\apj] {10.1086/164079}, \href
  {https://ui.adsabs.harvard.edu/abs/1986ApJ...303..336G} {303, 336}

\bibitem[\protect\citeauthoryear{{Giallongo} et~al.,}{{Giallongo}
  et~al.}{2019}]{gia19}
{Giallongo} E.,  et~al., 2019, \mn@doi [\apj] {10.3847/1538-4357/ab39e1}, \href
  {https://ui.adsabs.harvard.edu/abs/2019ApJ...884...19G} {884, 19}

\bibitem[\protect\citeauthoryear{{Grimm}, {Gilfanov}  \& {Sunyaev}}{{Grimm}
  et~al.}{2003}]{gri03}
{Grimm} H.~J.,  {Gilfanov} M.,   {Sunyaev} R.,  2003, \mn@doi [\mnras]
  {10.1046/j.1365-8711.2003.06224.x}, \href
  {https://ui.adsabs.harvard.edu/abs/2003MNRAS.339..793G} {339, 793}

\bibitem[\protect\citeauthoryear{{Grogin} et~al.,}{{Grogin}
  et~al.}{2011}]{gro11}
{Grogin} N.~A.,  et~al., 2011, \mn@doi [\apjs] {10.1088/0067-0049/197/2/35},
  \href {https://ui.adsabs.harvard.edu/abs/2011ApJS..197...35G} {197, 35}

\bibitem[\protect\citeauthoryear{{Guaita} et~al.,}{{Guaita}
  et~al.}{2017}]{gua17}
{Guaita} L.,  et~al., 2017, \mn@doi [\aap] {10.1051/0004-6361/201730603}, \href
  {https://ui.adsabs.harvard.edu/abs/2017A&A...606A..19G} {606, A19}

\bibitem[\protect\citeauthoryear{{Guo} et~al.,}{{Guo} et~al.}{2013}]{guo13}
{Guo} Y.,  et~al., 2013, \mn@doi [\apjs] {10.1088/0067-0049/207/2/24}, \href
  {https://ui.adsabs.harvard.edu/abs/2013ApJS..207...24G} {207, 24}

\bibitem[\protect\citeauthoryear{{Heger}, {Fryer}, {Woosley}, {Langer}  \&
  {Hartmann}}{{Heger} et~al.}{2003}]{heg03}
{Heger} A.,  {Fryer} C.~L.,  {Woosley} S.~E.,  {Langer} N.,   {Hartmann} D.~H.,
   2003, \mn@doi [\apj] {10.1086/375341}, \href
  {https://ui.adsabs.harvard.edu/abs/2003ApJ...591..288H} {591, 288}

\bibitem[\protect\citeauthoryear{{Hornschemeier}, {Heckman}, {Ptak}, {Tremonti}
   \& {Colbert}}{{Hornschemeier} et~al.}{2005}]{hor05}
{Hornschemeier} A.~E.,  {Heckman} T.~M.,  {Ptak} A.~F.,  {Tremonti} C.~A.,
  {Colbert} E.~J.~M.,  2005, \mn@doi [\aj] {10.1086/426371}, \href
  {https://ui.adsabs.harvard.edu/abs/2005AJ....129...86H} {129, 86}

\bibitem[\protect\citeauthoryear{Hunter}{Hunter}{2007}]{plt}
Hunter J.~D.,  2007, Computing In Science \& Engineering, 9, 90

\bibitem[\protect\citeauthoryear{{Iben}, {Tutukov}  \& {Yungelson}}{{Iben}
  et~al.}{1995}]{ibe95}
{Iben} Icko J.,  {Tutukov} A.~V.,   {Yungelson} L.~R.,  1995, \mn@doi [\apjs]
  {10.1086/192217}, \href
  {https://ui.adsabs.harvard.edu/abs/1995ApJS..100..217I} {100, 217}

\bibitem[\protect\citeauthoryear{{Kaaret}, {Schmitt}  \& {Gorski}}{{Kaaret}
  et~al.}{2011}]{kar11}
{Kaaret} P.,  {Schmitt} J.,   {Gorski} M.,  2011, \mn@doi [\apj]
  {10.1088/0004-637X/741/1/10}, \href
  {https://ui.adsabs.harvard.edu/abs/2011ApJ...741...10K} {741, 10}

\bibitem[\protect\citeauthoryear{{Kehrig}, {Guerrero}, {V{\'\i}lchez}  \&
  {Ramos-Larios}}{{Kehrig} et~al.}{2021}]{keh21}
{Kehrig} C.,  {Guerrero} M.~A.,  {V{\'\i}lchez} J.~M.,   {Ramos-Larios} G.,
  2021, \mn@doi [\apjl] {10.3847/2041-8213/abe41b}, \href
  {https://ui.adsabs.harvard.edu/abs/2021ApJ...908L..54K} {908, L54}

\bibitem[\protect\citeauthoryear{{Koekemoer} et~al.,}{{Koekemoer}
  et~al.}{2011}]{koe11}
{Koekemoer} A.~M.,  et~al., 2011, \mn@doi [\apjs] {10.1088/0067-0049/197/2/36},
  \href {https://ui.adsabs.harvard.edu/abs/2011ApJS..197...36K} {197, 36}

\bibitem[\protect\citeauthoryear{{Kouroumpatzakis} et~al.,}{{Kouroumpatzakis}
  et~al.}{2020}]{kou20}
{Kouroumpatzakis} K.,  et~al., 2020, \mn@doi [\mnras] {10.1093/mnras/staa1063},
  \href {https://ui.adsabs.harvard.edu/abs/2020MNRAS.494.5967K} {494, 5967}

\bibitem[\protect\citeauthoryear{{Kovlakas}, {Zezas}, {Andrews}, {Basu-Zych},
  {Fragos}, {Hornschemeier}, {Lehmer}  \& {Ptak}}{{Kovlakas}
  et~al.}{2020}]{kov20}
{Kovlakas} K.,  {Zezas} A.,  {Andrews} J.~J.,  {Basu-Zych} A.,  {Fragos} T.,
  {Hornschemeier} A.,  {Lehmer} B.,   {Ptak} A.,  2020, \mn@doi [\mnras]
  {10.1093/mnras/staa2481}, \href
  {https://ui.adsabs.harvard.edu/abs/2020MNRAS.498.4790K} {498, 4790}

\bibitem[\protect\citeauthoryear{{Kretschmar} et~al.,}{{Kretschmar}
  et~al.}{2021}]{kre21}
{Kretschmar} P.,  et~al., 2021, arXiv e-prints, \href
  {https://ui.adsabs.harvard.edu/abs/2021arXiv210413148K} {p. arXiv:2104.13148}

\bibitem[\protect\citeauthoryear{{Kroupa}}{{Kroupa}}{2001}]{kro01}
{Kroupa} P.,  2001, \mn@doi [\mnras] {10.1046/j.1365-8711.2001.04022.x}, \href
  {https://ui.adsabs.harvard.edu/abs/2001MNRAS.322..231K} {322, 231}

\bibitem[\protect\citeauthoryear{{Laporte}, {Nakajima}, {Ellis}, {Zitrin},
  {Stark}, {Mainali}  \& {Roberts-Borsani}}{{Laporte} et~al.}{2017}]{lap17}
{Laporte} N.,  {Nakajima} K.,  {Ellis} R.~S.,  {Zitrin} A.,  {Stark} D.~P.,
  {Mainali} R.,   {Roberts-Borsani} G.~W.,  2017, \mn@doi [\apj]
  {10.3847/1538-4357/aa96a8}, \href
  {https://ui.adsabs.harvard.edu/abs/2017ApJ...851...40L} {851, 40}

\bibitem[\protect\citeauthoryear{{Lehmer}, {Alexander}, {Bauer}, {Brandt},
  {Goulding}, {Jenkins}, {Ptak}  \& {Roberts}}{{Lehmer} et~al.}{2010}]{leh10}
{Lehmer} B.~D.,  {Alexander} D.~M.,  {Bauer} F.~E.,  {Brandt} W.~N.,
  {Goulding} A.~D.,  {Jenkins} L.~P.,  {Ptak} A.,   {Roberts} T.~P.,  2010,
  \mn@doi [\apj] {10.1088/0004-637X/724/1/559}, \href
  {https://ui.adsabs.harvard.edu/abs/2010ApJ...724..559L} {724, 559}

\bibitem[\protect\citeauthoryear{{Lehmer} et~al.,}{{Lehmer}
  et~al.}{2016}]{leh16}
{Lehmer} B.~D.,  et~al., 2016, \mn@doi [\apj] {10.3847/0004-637X/825/1/7},
  \href {https://ui.adsabs.harvard.edu/abs/2016ApJ...825....7L} {825, 7}

\bibitem[\protect\citeauthoryear{{Lehmer} et~al.,}{{Lehmer}
  et~al.}{2019}]{leh19}
{Lehmer} B.~D.,  et~al., 2019, \mn@doi [\apjs] {10.3847/1538-4365/ab22a8},
  \href {https://ui.adsabs.harvard.edu/abs/2019ApJS..243....3L} {243, 3}

\bibitem[\protect\citeauthoryear{{Lehmer} et~al.,}{{Lehmer}
  et~al.}{2021}]{leh21}
{Lehmer} B.~D.,  et~al., 2021, \mn@doi [\apj] {10.3847/1538-4357/abcec1}, \href
  {https://ui.adsabs.harvard.edu/abs/2021ApJ...907...17L} {907, 17}

\bibitem[\protect\citeauthoryear{{Leitherer}, {Ortiz Ot{\'a}lvaro}, {Bresolin},
  {Kudritzki}, {Lo Faro}, {Pauldrach}, {Pettini}  \& {Rix}}{{Leitherer}
  et~al.}{2010}]{lei10}
{Leitherer} C.,  {Ortiz Ot{\'a}lvaro} P.~A.,  {Bresolin} F.,  {Kudritzki}
  R.-P.,  {Lo Faro} B.,  {Pauldrach} A. W.~A.,  {Pettini} M.,   {Rix} S.~A.,
  2010, \mn@doi [\apjs] {10.1088/0067-0049/189/2/309}, \href
  {https://ui.adsabs.harvard.edu/abs/2010ApJS..189..309L} {189, 309}

\bibitem[\protect\citeauthoryear{{Linden}, {Kalogera}, {Sepinsky}, {Prestwich},
  {Zezas}  \& {Gallagher}}{{Linden} et~al.}{2010}]{lin10}
{Linden} T.,  {Kalogera} V.,  {Sepinsky} J.~F.,  {Prestwich} A.,  {Zezas} A.,
  {Gallagher} J.~S.,  2010, \mn@doi [\apj] {10.1088/0004-637X/725/2/1984},
  \href {https://ui.adsabs.harvard.edu/abs/2010ApJ...725.1984L} {725, 1984}

\bibitem[\protect\citeauthoryear{{Luo} et~al.,}{{Luo} et~al.}{2017}]{cdfs7ms}
{Luo} B.,  et~al., 2017, \mn@doi [\apjs] {10.3847/1538-4365/228/1/2}, \href
  {https://ui.adsabs.harvard.edu/abs/2017ApJS..228....2L} {228, 2}

\bibitem[\protect\citeauthoryear{{Madau} \& {Dickinson}}{{Madau} \&
  {Dickinson}}{2014}]{mad14}
{Madau} P.,  {Dickinson} M.,  2014, \mn@doi [\araa]
  {10.1146/annurev-astro-081811-125615}, \href
  {https://ui.adsabs.harvard.edu/abs/2014ARA&A..52..415M} {52, 415}

\bibitem[\protect\citeauthoryear{{Madau} \& {Fragos}}{{Madau} \&
  {Fragos}}{2017}]{mad17}
{Madau} P.,  {Fragos} T.,  2017, \mn@doi [\apj] {10.3847/1538-4357/aa6af9},
  \href {https://ui.adsabs.harvard.edu/abs/2017ApJ...840...39M} {840, 39}

\bibitem[\protect\citeauthoryear{{Magliocchetti} et~al.,}{{Magliocchetti}
  et~al.}{2020}]{mag20}
{Magliocchetti} M.,  et~al., 2020, \mn@doi [\mnras] {10.1093/mnras/staa410},
  \href {https://ui.adsabs.harvard.edu/abs/2020MNRAS.493.3838M} {493, 3838}

\bibitem[\protect\citeauthoryear{{Mainali} et~al.,}{{Mainali}
  et~al.}{2018}]{mai18}
{Mainali} R.,  et~al., 2018, \mn@doi [\mnras] {10.1093/mnras/sty1640}, \href
  {https://ui.adsabs.harvard.edu/abs/2018MNRAS.479.1180M} {479, 1180}

\bibitem[\protect\citeauthoryear{{Marchi} et~al.,}{{Marchi}
  et~al.}{2018}]{mar18}
{Marchi} F.,  et~al., 2018, \mn@doi [\aap] {10.1051/0004-6361/201732133}, \href
  {https://ui.adsabs.harvard.edu/abs/2018A&A...614A..11M} {614, A11}

\bibitem[\protect\citeauthoryear{{McLure} et~al.,}{{McLure}
  et~al.}{2018}]{mcl18}
{McLure} R.~J.,  et~al., 2018, \mn@doi [\mnras] {10.1093/mnras/sty1213}, \href
  {https://ui.adsabs.harvard.edu/abs/2018MNRAS.479...25M} {479, 25}

\bibitem[\protect\citeauthoryear{{Meiksin}, {Khochfar}, {Paardekooper}, {Dalla
  Vecchia}  \& {Kohn}}{{Meiksin} et~al.}{2017}]{mei17}
{Meiksin} A.,  {Khochfar} S.,  {Paardekooper} J.-P.,  {Dalla Vecchia} C.,
  {Kohn} S.,  2017, \mn@doi [\mnras] {10.1093/mnras/stx1857}, \href
  {https://ui.adsabs.harvard.edu/abs/2017MNRAS.471.3632M} {471, 3632}

\bibitem[\protect\citeauthoryear{{Mellema} et~al.,}{{Mellema}
  et~al.}{2013}]{mel13}
{Mellema} G.,  et~al., 2013, \mn@doi [Experimental Astronomy]
  {10.1007/s10686-013-9334-5}, \href
  {https://ui.adsabs.harvard.edu/abs/2013ExA....36..235M} {36, 235}

\bibitem[\protect\citeauthoryear{{Mesinger}, {Ferrara}  \&
  {Spiegel}}{{Mesinger} et~al.}{2013}]{mes13}
{Mesinger} A.,  {Ferrara} A.,   {Spiegel} D.~S.,  2013, \mn@doi [\mnras]
  {10.1093/mnras/stt198}, \href
  {https://ui.adsabs.harvard.edu/abs/2013MNRAS.431..621M} {431, 621}

\bibitem[\protect\citeauthoryear{{Mineo}, {Gilfanov}  \& {Sunyaev}}{{Mineo}
  et~al.}{2012}]{min12}
{Mineo} S.,  {Gilfanov} M.,   {Sunyaev} R.,  2012, \mn@doi [\mnras]
  {10.1111/j.1365-2966.2011.19862.x}, \href
  {https://ui.adsabs.harvard.edu/abs/2012MNRAS.419.2095M} {419, 2095}

\bibitem[\protect\citeauthoryear{{Naidu}, {Tacchella}, {Mason}, {Bose}, {Oesch}
   \& {Conroy}}{{Naidu} et~al.}{2020}]{nai20}
{Naidu} R.~P.,  {Tacchella} S.,  {Mason} C.~A.,  {Bose} S.,  {Oesch} P.~A.,
  {Conroy} C.,  2020, \mn@doi [\apj] {10.3847/1538-4357/ab7cc9}, \href
  {https://ui.adsabs.harvard.edu/abs/2020ApJ...892..109N} {892, 109}

\bibitem[\protect\citeauthoryear{{Nakajima}, {Fletcher}, {Ellis}, {Robertson}
  \& {Iwata}}{{Nakajima} et~al.}{2018}]{nak18}
{Nakajima} K.,  {Fletcher} T.,  {Ellis} R.~S.,  {Robertson} B.~E.,   {Iwata}
  I.,  2018, \mn@doi [\mnras] {10.1093/mnras/sty750}, \href
  {https://ui.adsabs.harvard.edu/abs/2018MNRAS.477.2098N} {477, 2098}

\bibitem[\protect\citeauthoryear{{Oke} \& {Gunn}}{{Oke} \&
  {Gunn}}{1983}]{oke83}
{Oke} J.~B.,  {Gunn} J.~E.,  1983, \mn@doi [\apj] {10.1086/160817}, \href
  {https://ui.adsabs.harvard.edu/abs/1983ApJ...266..713O} {266, 713}

\bibitem[\protect\citeauthoryear{{Ouchi}, {Ono}  \& {Shibuya}}{{Ouchi}
  et~al.}{2020}]{ouc20}
{Ouchi} M.,  {Ono} Y.,   {Shibuya} T.,  2020, \mn@doi [\araa]
  {10.1146/annurev-astro-032620-021859}, \href
  {https://ui.adsabs.harvard.edu/abs/2020ARA&A..58..617O} {58, 617}

\bibitem[\protect\citeauthoryear{{Pacucci}, {Mesinger}, {Mineo}  \&
  {Ferrara}}{{Pacucci} et~al.}{2014}]{pac14}
{Pacucci} F.,  {Mesinger} A.,  {Mineo} S.,   {Ferrara} A.,  2014, \mn@doi
  [\mnras] {10.1093/mnras/stu1240}, \href
  {https://ui.adsabs.harvard.edu/abs/2014MNRAS.443..678P} {443, 678}

\bibitem[\protect\citeauthoryear{{Pentericci} et~al.,}{{Pentericci}
  et~al.}{2018}]{pen18}
{Pentericci} L.,  et~al., 2018, \mn@doi [\aap] {10.1051/0004-6361/201833047},
  \href {https://ui.adsabs.harvard.edu/abs/2018A&A...616A.174P} {616, A174}

\bibitem[\protect\citeauthoryear{P\'erez \& Granger}{P\'erez \&
  Granger}{2007}]{ipython}
P\'erez F.,  Granger B.~E.,  2007, \mn@doi [Computing in Science and
  Engineering] {10.1109/MCSE.2007.53}, 9, 21

\bibitem[\protect\citeauthoryear{{Planck Collaboration} et~al.,}{{Planck
  Collaboration} et~al.}{2016}]{planck}
{Planck Collaboration} et~al., 2016, \mn@doi [\aap]
  {10.1051/0004-6361/201525830}, \href
  {https://ui.adsabs.harvard.edu/abs/2016A&A...594A..13P} {594, A13}

\bibitem[\protect\citeauthoryear{{Plat}, {Charlot}, {Bruzual}, {Feltre},
  {Vidal-Garc{\'\i}a}, {Morisset}, {Chevallard}  \& {Todt}}{{Plat}
  et~al.}{2019}]{pla19}
{Plat} A.,  {Charlot} S.,  {Bruzual} G.,  {Feltre} A.,  {Vidal-Garc{\'\i}a} A.,
   {Morisset} C.,  {Chevallard} J.,   {Todt} H.,  2019, \mn@doi [\mnras]
  {10.1093/mnras/stz2616}, \href
  {https://ui.adsabs.harvard.edu/abs/2019MNRAS.490..978P} {490, 978}

\bibitem[\protect\citeauthoryear{{Ponnada}, {Brorby}  \& {Kaaret}}{{Ponnada}
  et~al.}{2020}]{pon20}
{Ponnada} S.,  {Brorby} M.,   {Kaaret} P.,  2020, \mn@doi [\mnras]
  {10.1093/mnras/stz2929}, \href
  {https://ui.adsabs.harvard.edu/abs/2020MNRAS.491.3606P} {491, 3606}

\bibitem[\protect\citeauthoryear{{Prestwich}, {Tsantaki}, {Zezas}, {Jackson},
  {Roberts}, {Foltz}, {Linden}  \& {Kalogera}}{{Prestwich}
  et~al.}{2013}]{pre13}
{Prestwich} A.~H.,  {Tsantaki} M.,  {Zezas} A.,  {Jackson} F.,  {Roberts}
  T.~P.,  {Foltz} R.,  {Linden} T.,   {Kalogera} V.,  2013, \mn@doi [\apj]
  {10.1088/0004-637X/769/2/92}, \href
  {https://ui.adsabs.harvard.edu/abs/2013ApJ...769...92P} {769, 92}

\bibitem[\protect\citeauthoryear{{Ranalli}, {Comastri}  \& {Setti}}{{Ranalli}
  et~al.}{2003}]{ran03}
{Ranalli} P.,  {Comastri} A.,   {Setti} G.,  2003, \mn@doi [\aap]
  {10.1051/0004-6361:20021600}, \href
  {https://ui.adsabs.harvard.edu/abs/2003A&A...399...39R} {399, 39}

\bibitem[\protect\citeauthoryear{{Rappaport}, {Podsiadlowski}  \&
  {Pfahl}}{{Rappaport} et~al.}{2005}]{rap05}
{Rappaport} S.~A.,  {Podsiadlowski} P.,   {Pfahl} E.,  2005, \mn@doi [\mnras]
  {10.1111/j.1365-2966.2004.08489.x}, \href
  {https://ui.adsabs.harvard.edu/abs/2005MNRAS.356..401R} {356, 401}

\bibitem[\protect\citeauthoryear{{Robertson} et~al.,}{{Robertson}
  et~al.}{2013}]{rob13}
{Robertson} B.~E.,  et~al., 2013, \mn@doi [\apj] {10.1088/0004-637X/768/1/71},
  \href {https://ui.adsabs.harvard.edu/abs/2013ApJ...768...71R} {768, 71}

\bibitem[\protect\citeauthoryear{{Robertson}, {Ellis}, {Furlanetto}  \&
  {Dunlop}}{{Robertson} et~al.}{2015}]{rob15}
{Robertson} B.~E.,  {Ellis} R.~S.,  {Furlanetto} S.~R.,   {Dunlop} J.~S.,
  2015, \mn@doi [\apjl] {10.1088/2041-8205/802/2/L19}, \href
  {https://ui.adsabs.harvard.edu/abs/2015ApJ...802L..19R} {802, L19}

\bibitem[\protect\citeauthoryear{{Saxena} et~al.,}{{Saxena}
  et~al.}{2020a}]{sax20b}
{Saxena} A.,  et~al., 2020a, \mn@doi [\mnras] {10.1093/mnras/staa1805}, \href
  {https://ui.adsabs.harvard.edu/abs/2020MNRAS.496.3796S} {496, 3796}

\bibitem[\protect\citeauthoryear{{Saxena} et~al.,}{{Saxena}
  et~al.}{2020b}]{sax20}
{Saxena} A.,  et~al., 2020b, \mn@doi [\aap] {10.1051/0004-6361/201937170},
  \href {https://ui.adsabs.harvard.edu/abs/2020A&A...636A..47S} {636, A47}

\bibitem[\protect\citeauthoryear{{Schaerer}, {Fragos}  \& {Izotov}}{{Schaerer}
  et~al.}{2019}]{sch19}
{Schaerer} D.,  {Fragos} T.,   {Izotov} Y.~I.,  2019, \mn@doi [\aap]
  {10.1051/0004-6361/201935005}, \href
  {https://ui.adsabs.harvard.edu/abs/2019A&A...622L..10S} {622, L10}

\bibitem[\protect\citeauthoryear{{Senchyna}, {Stark}, {Mirocha}, {Reines},
  {Charlot}, {Jones}  \& {Mulchaey}}{{Senchyna} et~al.}{2020}]{sen20}
{Senchyna} P.,  {Stark} D.~P.,  {Mirocha} J.,  {Reines} A.~E.,  {Charlot} S.,
  {Jones} T.,   {Mulchaey} J.~S.,  2020, \mn@doi [\mnras]
  {10.1093/mnras/staa586}, \href
  {https://ui.adsabs.harvard.edu/abs/2020MNRAS.494..941S} {494, 941}

\bibitem[\protect\citeauthoryear{{Shapley}, {Steidel}, {Pettini}  \&
  {Adelberger}}{{Shapley} et~al.}{2003}]{sha03}
{Shapley} A.~E.,  {Steidel} C.~C.,  {Pettini} M.,   {Adelberger} K.~L.,  2003,
  \mn@doi [\apj] {10.1086/373922}, \href
  {https://ui.adsabs.harvard.edu/abs/2003ApJ...588...65S} {588, 65}

\bibitem[\protect\citeauthoryear{{Stark}}{{Stark}}{2016}]{star16}
{Stark} D.~P.,  2016, \mn@doi [\araa] {10.1146/annurev-astro-081915-023417},
  \href {https://ui.adsabs.harvard.edu/abs/2016ARA&A..54..761S} {54, 761}

\bibitem[\protect\citeauthoryear{{Stark} et~al.,}{{Stark}
  et~al.}{2017}]{star17}
{Stark} D.~P.,  et~al., 2017, \mn@doi [\mnras] {10.1093/mnras/stw2233}, \href
  {https://ui.adsabs.harvard.edu/abs/2017MNRAS.464..469S} {464, 469}

\bibitem[\protect\citeauthoryear{{Tauris} \& {van den Heuvel}}{{Tauris} \& {van
  den Heuvel}}{2006}]{tau06}
{Tauris} T.~M.,  {van den Heuvel} E.~P.~J.,  2006, {Formation and evolution of
  compact stellar X-ray sources}.
pp 623--665

\bibitem[\protect\citeauthoryear{{Taylor}}{{Taylor}}{2005}]{topcat}
{Taylor} M.~B.,  2005, in {Shopbell} P.,  {Britton} M.,   {Ebert} R.,  eds,
  Astronomical Society of the Pacific Conference Series Vol. 347, Astronomical
  Data Analysis Software and Systems XIV. p.~29

\bibitem[\protect\citeauthoryear{{Verhamme} et~al.,}{{Verhamme}
  et~al.}{2018}]{ver18}
{Verhamme} A.,  et~al., 2018, \mn@doi [\mnras] {10.1093/mnrasl/sly058}, \href
  {https://ui.adsabs.harvard.edu/abs/2018MNRAS.478L..60V} {478, L60}

\bibitem[\protect\citeauthoryear{{Vito} et~al.,}{{Vito} et~al.}{2018}]{vit18}
{Vito} F.,  et~al., 2018, \mn@doi [\mnras] {10.1093/mnras/stx2486}, \href
  {https://ui.adsabs.harvard.edu/abs/2018MNRAS.473.2378V} {473, 2378}

\bibitem[\protect\citeauthoryear{{Warszawski}, {Geil}  \&
  {Wyithe}}{{Warszawski} et~al.}{2009}]{war09}
{Warszawski} L.,  {Geil} P.~M.,   {Wyithe} J.~S.~B.,  2009, \mn@doi [\mnras]
  {10.1111/j.1365-2966.2009.14781.x}, \href
  {https://ui.adsabs.harvard.edu/abs/2009MNRAS.396.1106W} {396, 1106}

\bibitem[\protect\citeauthoryear{pandas~development team}{pandas~development
  team}{2020}]{pandas}
pandas~development team T.,  2020, pandas-dev/pandas: Pandas 1.1.5,
  \mn@doi{10.5281/zenodo.4309786}, \url
  {https://doi.org/10.5281/zenodo.4309786}

\bibitem[\protect\citeauthoryear{{van de Voort}, {Schaye}, {Altay}  \&
  {Theuns}}{{van de Voort} et~al.}{2012}]{voo12}
{van de Voort} F.,  {Schaye} J.,  {Altay} G.,   {Theuns} T.,  2012, \mn@doi
  [\mnras] {10.1111/j.1365-2966.2012.20487.x}, \href
  {https://ui.adsabs.harvard.edu/abs/2012MNRAS.421.2809V} {421, 2809}

\makeatother
\end{thebibliography}





\bsp	
\label{lastpage}
\end{document}